\documentclass[
  a4paper,
  superscriptaddress,
  twocolumn,
  prl,
  preprintnumbers,
  floatfix,
  secnumarabic,
  amssymb
]
{revtex4-2}
\usepackage{lineno}
\usepackage{graphicx}  
\usepackage{dcolumn}   
\usepackage{bm}        
\usepackage{amssymb}   
\usepackage{amsmath,stmaryrd}
\usepackage{blkarray, multirow, graphicx, diagbox, color, xcolor, colortbl}
\usepackage{bbm, bbold}
\usepackage{glossaries}
\usepackage{hyphenat}
\usepackage{ifthen}
\usepackage{xkeyval}
\usepackage{moreverb}
\usepackage{rotating}
\usepackage{wrapfig}
\usepackage{slashbox}
\usepackage{xspace}
\usepackage{nicefrac}
\usepackage[]{units}
\usepackage[]{natbib}
\usepackage{physics}
\usepackage{booktabs}
\usepackage{braket}
\usepackage[inline]{enumitem}
\usepackage{tabto}
\usepackage{listings}
\usepackage{xstring}
\usepackage{lipsum}
\usepackage{scalerel}

\def\ReplaceStr#1{%
	\IfSubStr{#1}{p}{%
		\StrSubstitute{#1}{p}{.}}{#1}}

\usepackage[caption=false]{subfig}
\newcommand\subfigref[1]{\protect\subref{#1}}
\def\figurepath{./}

\usepackage[customcolors]{hf-tikz} 
\usepackage{tikz}
\usepackage{calc}
\usetikzlibrary{external}
\usepackage{pgffor}
\usepackage{pgfplots}
\pgfplotsset{compat=1.13}
\usepackage{pgfplotstable}
\usepgfplotslibrary{groupplots}
\usepgfplotslibrary{fillbetween}
\usepgfplotslibrary{colorbrewer}

\tikzstyle{n} = [draw,shape=ellipse,minimum size=1.5em,inner sep=0pt,fill=white!20, minimum width=2.5em]
\tikzstyle{Init} = [n,color=green,fill=green!20,text=black]
\tikzstyle{Fin} = [n,color=red,fill=red!20,text=black]
\tikzstyle{Ghost} = [minimum size=1.5em,inner sep=0pt,color=white,text=black]
\tikzstyle{Multiple} = [draw,shape=rect,minimum size=2em,inner sep=0pt]

\tikzstyle{ghostA} = [text=red!70,thick, minimum size=2*(5pt-\pgflinewidth), inner sep=0pt, outer sep=0pt]
\tikzstyle{ghostB} = [text=blue!70,thick, minimum size=2*(5pt-\pgflinewidth), inner sep=0pt, outer sep=0pt]
\tikzstyle{siteA} = [regular polygon, regular polygon sides=3, shape border rotate= 30, draw=red!50,fill=red!20,thick,inner sep=0pt,minimum width=1.5em,font=\footnotesize]
\tikzstyle{siteB} = [regular polygon, regular polygon sides=3, shape border rotate= -30, draw=green!50,fill=green!20,thick,inner sep=0pt,minimum width=1.5em,font=\footnotesize]
\tikzstyle{op} = [regular polygon, regular polygon sides=4, draw=orange!50, fill=orange!20, thick, inner sep=0.2pt, minimum width=0.25em, minimum height=0.5em,font=\footnotesize]
\tikzstyle{gate} = [rectangle, rounded corners=2pt, draw=orange!50, fill=orange!20, thick, inner sep=0.2pt, minimum width=0.25em, minimum height=0.5em,font=\footnotesize]
\tikzstyle{opghost} = [regular polygon, regular polygon sides=4, thick, inner sep=0.2pt, minimum width=1.25em, minimum height=1.5em,font=\footnotesize]
\tikzstyle{site} = [circle,draw=blue!50,fill=blue!20,thick,inner sep=0.2pt,minimum width=1.25em,font=\footnotesize]
\tikzstyle{hiddensite} = [circle,draw=white!50,fill=white!20,thick,inner sep=0.2pt,minimum width=1.25em,font=\footnotesize]
\tikzstyle{nosite} = [circle,draw=white,fill=white,thick,inner sep=0.1pt,minimum width=1.5em]
\tikzstyle{ghost} = [font=\footnotesize]
\tikzstyle{intersite} = [regular polygon, regular polygon sides=4, shape border rotate= 45, draw=black!50,fill=black!20,thick,inner sep=0pt,minimum width=1.5em]
\tikzstyle{ld} = [inner sep=1pt, font=\small]
\tikzstyle{unsite} = [circle, outer sep=0pt,inner sep=0.2pt,minimum width=1.25em]

\definecolor{colorA}{rgb} {0.58,0,0.8275}
\definecolor{colorB}{rgb} {0.11,0.663,0.51}
\definecolor{colorC}{rgb} {0.3373,0.7059,0.9137}
\definecolor{colorD}{rgb} {0.902,0.6235,0}
\definecolor{colorE}{rgb} {0.9451,0.902,0.3255}
\definecolor{colorF}{rgb} {0.3373,0.3255,0.902}
\definecolor{colorG}{rgb} {0.9451,0.3255,0.3373}
\definecolor{cbColorA}{HTML} {2D4C8D}
\definecolor{cbColorB}{HTML} {EB821D}
\definecolor{cbColorC}{HTML} {C3342F}
\definecolor{cbColorD}{HTML} {24451B}
\definecolor{cbColorE}{HTML} {2C0041}
\definecolor{cbColorF}{HTML} {D9D9D9}

\definecolor{stBlue}{HTML} {5566AA}
\definecolor{stGreen}{HTML} {117733}
\definecolor{stCyan}{HTML} {33BBEE}
\definecolor{stTeal}{HTML} {009988}
\definecolor{stOrange}{HTML} {EE7733}
\definecolor{stYellow}{HTML} {F7F056}
\definecolor{stRed}{HTML} {CC3311}
\definecolor{stMagenta}{HTML} {EE3377}
\definecolor{stGrey}{HTML} {BBBBBB}

\usetikzlibrary
{
	calc,
	decorations,
	plotmarks,
	patterns,
	positioning,
	petri,
	arrows,
	decorations.markings,
	backgrounds,
	fit,
	graphs,
	shapes.geometric,
	decorations.pathmorphing,
	shapes.misc,
	shapes,
	tikzmark,
	pgfplots.colorbrewer,
	fpu,
	svg.path
}

\usepackage{ocgx}

\usetikzlibrary{ocgx}

\pgfplotsset{
        cycle from colormap manual style/.style={
            x=3cm,y=10pt,ytick=\empty,
            stack plots=y,
            every axis plot/.style={line width=2pt},
        },
}

\tikzset{>=stealth}

\tikzset{->-/.style={decoration={
			markings,
			mark=at position .5 with {\arrow{>}}},postaction={decorate}}}

\tikzset{-<-/.style={decoration={
			markings,
			mark=at position .5 with {\arrow{<}}},postaction={decorate}}}

\tikzstyle{orientedsnake} = [
decorate, 
decoration={snake},
->
]  
\tikzstyle{orientedshortarrow} = [
decoration={markings,
	mark=at position .33 with {\arrow{>}}},
postaction={decorate}
]  
\tikzstyle{orientedlongarrow} = [
decoration={markings,
	mark=at position .67 with {\arrow{>}}},
postaction={decorate}
]
\tikzset{dbl/.style={double,
		double equal sign distance,
		-implies,
		shorten >=10pt,
		shorten <=10pt}}
\tikzset{
	between/.style args={#1 and #2}{
		at = ($(#1)!0.5!(#2)$)
	}
}

\tikzstyle{process} = [rectangle, minimum width=3cm, minimum height=1cm, text centered, text width=5cm, draw=black]
\tikzstyle{io} = [trapezium, trapezium left angle=70, trapezium right angle=110, minimum width=3cm, minimum height=1cm, text centered, text width=7cm, draw=black]
\tikzstyle{choose} = [diamond, inner sep=1pt, minimum width=2cm, minimum height=2cm, text centered, text width=1.5cm, draw=black]
\tikzstyle{arrow} =[thick,->, >=stealth]

\pgfmathdeclarefunction{peierlspotential}{6}%
{
	\pgfmathparse
	{
		#2 / (#3 * #5) 
		* sin(deg(#1 * #3 - #3 * #5 * (x)))
		* exp(-( ( #1 - #5 * ((x) - #4) ) )^2.0 / ( #6^2.0 ) )
	}%
}

\pgfmathdeclarefunction{linearFct}{2}%
{%
	\pgfmathparse{#1*x+#2}%
}
\pgfmathdeclarefunction{quadFct}{3}%
{%
	\pgfmathparse{#1*x*x+#2*x+#3}%
}
\pgfmathdeclarefunction{logFct}{2}%
{%
	\pgfmathparse{#1*log10(x)+#2}%
}
\pgfmathdeclarefunction{expFct}{2}%
{%
	\pgfmathparse{#1*exp(-x/#2)}%
}

\newboolean{buildCurrentBlockInline}
\setboolean{buildCurrentBlockInline}{false}
\newboolean{rebuildCurrentBlock}
\setboolean{rebuildCurrentBlock}{false}

\newboolean{rebuildData}
\setboolean{rebuildData}{false}

\newboolean{removeData}
\setboolean{removeData}{false}

\graphicspath{{figures/}}

\newboolean{buildtikzpics}
\setboolean{buildtikzpics}{false}
\newif\ifrebuildtikz
\newif\ifChangeMode
\ChangeModetrue
\ChangeModefalse
\ifthenelse{\boolean{buildtikzpics}}
{
	\rebuildtikztrue
	\usetikzlibrary{external}
	\tikzexternalize[optimize=false,prefix=figures/autogen/]%
}
{
	\rebuildtikzfalse
}

\usepackage[normalem]{ulem}
\usepackage[colorlinks,bookmarks=false,citecolor=blue,linkcolor=black,urlcolor=blue]{hyperref}
\usepackage{cleveref}
\Crefname{appendix}{Appendix}{Appendices}
\Crefname{equation}{Equation}{Equations}
\Crefname{figure}{Figure}{Figures}
\Crefname{section}{Section}{Sections}
\Crefname{tabular}{Tabular}{Tabulars}
\crefname{appendix}{App.}{Apps.}
\crefname{equation}{Eq.}{Eqs.}
\crefname{figure}{Fig.}{Figs.}
\crefname{section}{Sec.}{Secs.}
\crefname{tabular}{Tab.}{Tabs.}

\newcommand{\symmps}{\textsc{SymMPS}}
\newcommand{\syten}{\textsc{SyTen}}

\ExplSyntaxOn
\DeclareExpandableDocumentCommand \eval { m } { \fp_eval:n { #1 } }
\ExplSyntaxOff

\makeatletter
\def\pgfplotsutil@decstringcounter#1{%
 \begingroup
  \c@pgf@counta=#1\relax
  \advance\c@pgf@counta by -1
  \edef#1{\the\c@pgf@counta}%
  \pgfmath@smuggleone#1%
 \endgroup
}%

\pgfplotsset{
/pgfplots/each nth point*/.style 2 args={%
/pgfplots/x filter/.append code={%
 \ifnum\coordindex=0
  \def\c@pgfplots@eachnthpoint@xfilter{0}%
  \edef\c@pgfplots@eachnthpoint@xfilter@cmp{#1}%
 \else
  \ifnum\coordindex>#2\relax
   \pgfplotsutil@advancestringcounter\c@pgfplots@eachnthpoint@xfilter
   \ifx\c@pgfplots@eachnthpoint@xfilter@cmp\c@pgfplots@eachnthpoint@xfilter
    \def\c@pgfplots@eachnthpoint@xfilter{0}%
   \else
    \let\pgfmathresult\pgfutil@empty
   \fi
  \fi
 \fi
}%
},
}
\makeatother

\newcommand{\printpgfnumberorder}[1]%
{%
	\pgfmathfloatparsenumber{#1}%
	\pgfmathfloattomacro{\pgfmathresult}{\Fn}{\Mn}{\En}%
	\pgfmathparse{\Fn==2 ? "-" : ""}%
	\edef\Sn{\pgfmathresult}%
	\Sn 10^{\En}%
}

\newcounter{marknumber}
\pgfplotsset{
	error bars/every nth mark/.style={
		/pgfplots/error bars/draw error bar/.prefix code={
			\pgfmathtruncatemacro\marknumbercheck{mod(floor(\themarknumber/2),#1)}
			\ifnum\marknumbercheck=0
			\else
			\begin{scope}[opacity=0]
				\fi
			},
			/pgfplots/error bars/draw error bar/.append code={
				\ifnum\marknumbercheck=0
				\else
			\end{scope}
			\fi
			\stepcounter{marknumber}    
		}
	}
}

\pgfplotsset{
    /pgf/declare function={
        vk(\x,\a,\b,\c) = \c*sqrt(1.0-(-2*cos(\x)-\b)/sqrt((-2*cos(\x)-\b)^2+\a^2));
    },
}

\pgfplotsset{
/pgfplots/colormap={magma}{
	rgb255=(0,0,4) rgb255=(0,0,6) rgb255=(1,0,7) rgb255=(1,1,9) rgb255=(1,1,11) rgb255=(2,2,13) rgb255=(2,2,15) rgb255=(3,3,17) rgb255=(4,3,19) rgb255=(4,4,21) rgb255=(5,4,23) rgb255=(6,5,25) rgb255=(7,5,27) rgb255=(8,6,29) rgb255=(9,7,31) rgb255=(10,7,34) rgb255=(11,8,36) rgb255=(12,9,38) rgb255=(13,10,40) rgb255=(14,10,42) rgb255=(15,11,44) rgb255=(16,12,47) rgb255=(17,12,49) rgb255=(18,13,51) rgb255=(20,13,53) rgb255=(21,14,56) rgb255=(22,14,58) rgb255=(23,15,60) rgb255=(24,15,63) rgb255=(26,16,65) rgb255=(27,16,68) rgb255=(28,16,70) rgb255=(30,16,73) rgb255=(31,17,75) rgb255=(32,17,77) rgb255=(34,17,80) rgb255=(35,17,82) rgb255=(37,17,85) rgb255=(38,17,87) rgb255=(40,17,89) rgb255=(42,17,92) rgb255=(43,17,94) rgb255=(45,16,96) rgb255=(47,16,98) rgb255=(48,16,101) rgb255=(50,16,103) rgb255=(52,16,104) rgb255=(53,15,106) rgb255=(55,15,108) rgb255=(57,15,110) rgb255=(59,15,111) rgb255=(60,15,113) rgb255=(62,15,114) rgb255=(64,15,115) rgb255=(66,15,116) rgb255=(67,15,117) rgb255=(69,15,118) rgb255=(71,15,119) rgb255=(72,16,120) rgb255=(74,16,121) rgb255=(75,16,121) rgb255=(77,17,122) rgb255=(79,17,123) rgb255=(80,18,123) rgb255=(82,18,124) rgb255=(83,19,124) rgb255=(85,19,125) rgb255=(87,20,125) rgb255=(88,21,126) rgb255=(90,21,126) rgb255=(91,22,126) rgb255=(93,23,126) rgb255=(94,23,127) rgb255=(96,24,127) rgb255=(97,24,127) rgb255=(99,25,127) rgb255=(101,26,128) rgb255=(102,26,128) rgb255=(104,27,128) rgb255=(105,28,128) rgb255=(107,28,128) rgb255=(108,29,128) rgb255=(110,30,129) rgb255=(111,30,129) rgb255=(113,31,129) rgb255=(115,31,129) rgb255=(116,32,129) rgb255=(118,33,129) rgb255=(119,33,129) rgb255=(121,34,129) rgb255=(122,34,129) rgb255=(124,35,129) rgb255=(126,36,129) rgb255=(127,36,129) rgb255=(129,37,129) rgb255=(130,37,129) rgb255=(132,38,129) rgb255=(133,38,129) rgb255=(135,39,129) rgb255=(137,40,129) rgb255=(138,40,129) rgb255=(140,41,128) rgb255=(141,41,128) rgb255=(143,42,128) rgb255=(145,42,128) rgb255=(146,43,128) rgb255=(148,43,128) rgb255=(149,44,128) rgb255=(151,44,127) rgb255=(153,45,127) rgb255=(154,45,127) rgb255=(156,46,127) rgb255=(158,46,126) rgb255=(159,47,126) rgb255=(161,47,126) rgb255=(163,48,126) rgb255=(164,48,125) rgb255=(166,49,125) rgb255=(167,49,125) rgb255=(169,50,124) rgb255=(171,51,124) rgb255=(172,51,123) rgb255=(174,52,123) rgb255=(176,52,123) rgb255=(177,53,122) rgb255=(179,53,122) rgb255=(181,54,121) rgb255=(182,54,121) rgb255=(184,55,120) rgb255=(185,55,120) rgb255=(187,56,119) rgb255=(189,57,119) rgb255=(190,57,118) rgb255=(192,58,117) rgb255=(194,58,117) rgb255=(195,59,116) rgb255=(197,60,116) rgb255=(198,60,115) rgb255=(200,61,114) rgb255=(202,62,114) rgb255=(203,62,113) rgb255=(205,63,112) rgb255=(206,64,112) rgb255=(208,65,111) rgb255=(209,66,110) rgb255=(211,66,109) rgb255=(212,67,109) rgb255=(214,68,108) rgb255=(215,69,107) rgb255=(217,70,106) rgb255=(218,71,105) rgb255=(220,72,105) rgb255=(221,73,104) rgb255=(222,74,103) rgb255=(224,75,102) rgb255=(225,76,102) rgb255=(226,77,101) rgb255=(228,78,100) rgb255=(229,80,99) rgb255=(230,81,98) rgb255=(231,82,98) rgb255=(232,84,97) rgb255=(234,85,96) rgb255=(235,86,96) rgb255=(236,88,95) rgb255=(237,89,95) rgb255=(238,91,94) rgb255=(238,93,93) rgb255=(239,94,93) rgb255=(240,96,93) rgb255=(241,97,92) rgb255=(242,99,92) rgb255=(243,101,92) rgb255=(243,103,91) rgb255=(244,104,91) rgb255=(245,106,91) rgb255=(245,108,91) rgb255=(246,110,91) rgb255=(246,112,91) rgb255=(247,113,91) rgb255=(247,115,92) rgb255=(248,117,92) rgb255=(248,119,92) rgb255=(249,121,92) rgb255=(249,123,93) rgb255=(249,125,93) rgb255=(250,127,94) rgb255=(250,128,94) rgb255=(250,130,95) rgb255=(251,132,96) rgb255=(251,134,96) rgb255=(251,136,97) rgb255=(251,138,98) rgb255=(252,140,99) rgb255=(252,142,99) rgb255=(252,144,100) rgb255=(252,146,101) rgb255=(252,147,102) rgb255=(253,149,103) rgb255=(253,151,104) rgb255=(253,153,105) rgb255=(253,155,106) rgb255=(253,157,107) rgb255=(253,159,108) rgb255=(253,161,110) rgb255=(253,162,111) rgb255=(253,164,112) rgb255=(254,166,113) rgb255=(254,168,115) rgb255=(254,170,116) rgb255=(254,172,117) rgb255=(254,174,118) rgb255=(254,175,120) rgb255=(254,177,121) rgb255=(254,179,123) rgb255=(254,181,124) rgb255=(254,183,125) rgb255=(254,185,127) rgb255=(254,187,128) rgb255=(254,188,130) rgb255=(254,190,131) rgb255=(254,192,133) rgb255=(254,194,134) rgb255=(254,196,136) rgb255=(254,198,137) rgb255=(254,199,139) rgb255=(254,201,141) rgb255=(254,203,142) rgb255=(253,205,144) rgb255=(253,207,146) rgb255=(253,209,147) rgb255=(253,210,149) rgb255=(253,212,151) rgb255=(253,214,152) rgb255=(253,216,154) rgb255=(253,218,156) rgb255=(253,220,157) rgb255=(253,221,159) rgb255=(253,223,161) rgb255=(253,225,163) rgb255=(252,227,165) rgb255=(252,229,166) rgb255=(252,230,168) rgb255=(252,232,170) rgb255=(252,234,172) rgb255=(252,236,174) rgb255=(252,238,176) rgb255=(252,240,177) rgb255=(252,241,179) rgb255=(252,243,181) rgb255=(252,245,183) rgb255=(251,247,185) rgb255=(251,249,187) rgb255=(251,250,189) rgb255=(251,252,191)}
}
\newacronym[shortplural={MPS}]{MPS}{MPS}{matrix\hyp product state}
\newacronym{MPO}{MPO}{matrix-product operator}
\newacronym{SVD}{SVD}{singular-value decomposition}
\newacronym{QCS}{QCS}{quantum-computer simulator}
\newacronym{FSM}{FSM}{finite-state machine}
\newacronym{ACA}{ACA}{adaptive cross approximation}
\newacronym{1D}{1D}{one\hyp dimensional}
\newacronym{QC}{QC}{quantum computer}
\newacronym{CDW}{CDW}{charge\hyp density wave}
\newacronym{SDW}{SDW}{spin\hyp density wave}
\newacronym{ARPES}{ARPES}{angle-resolved photoemission spectroscopy}
\newacronym{OBC}{OBC}{open-boundary conditions}
\newacronym{PBC}{PBC}{periodic-boundary conditions}
\newacronym{TEBD}{TEBD}{time-evolution block-decimation}
\newacronym{TDVP}{TDVP}{time\hyp dependent variational principle}
\newacronym{iff}{iff}{if and only if}
\newacronym{DFT}{DFT}{density\hyp functional theory}
\newacronym{DMFT}{DMFT}{dynamical mean\hyp field theory}
\newacronym{DMRG}{DMRG}{density\hyp matrix renormalization group}
\newacronym{ppDMRG}{ppDMRG}{projected purified density\hyp matrix renormalization group}
\newacronym{1DMRG}{1DMRG}{single-site density\hyp matrix renormalization group}
\newacronym{2DMRG}{2DMRG}{two-site density\hyp matrix renormalization group}
\newacronym{DMRG3S}{DMRG3S}{strictly single-site density\hyp matrix renormalization group}
\newacronym{iDMRG}{iDMRG}{inifinite\hyp size density\hyp matrix renormalization group}
\newacronym{tDMRG}{tDMRG}{time\hyp dependent density\hyp matrix renormalization group}
\newacronym{QMC}{QMC}{quantum Monte Carlo}
\newacronym{AIM}{AIM}{Anderson impurity model}
\newacronym{SIAM}{SIAM}{single impurity Anderson model}
\newacronym{LDA}{LDA}{local\hyp density approximation}
\newacronym{LBNL}{LBNL}{Lawrence Berkeley National Laboratory}
\newacronym{VQE}{VQE}{variational\hyp quantum eigensolver}
\newacronym{ED}{ED}{exact diagonalization}
\newacronym{QPT}{QPT}{quantum phase transition}
\newacronym{QCP}{QCP}{quantum critical point}
\newacronym{ETH}{ETH}{eigenstate thermalization hypothesis}
\newacronym{EHM}{EHM}{extended Hubbard model}
\newacronym{AKLT}{AKLT}{Affleck\hyp Lieb\hyp Kennedy\hyp Tasaki}
\newglossaryentry{QR}{name={QR},description={QR decomposition}}
\newacronym{TN}{TN}{tensor\hyp network}
\newacronym{TNS}{TNS}{tensor\hyp network state}
\newacronym{SM}{SM}{supplemental material}
\newacronym{NOO}{NOO}{natural orbital occupation}
\newacronym{NO}{NO}{natural orbital}
\newacronym{LRO}{LRO}{long\hyp range order}
\newacronym{qLRO}{qLRO}{quasi\hyp long\hyp range order}
\newacronym{SC}{SC}{Superconductivity}
\newacronym{VBGS}{VBGS}{valence bond ground-state}
\newacronym{PEPS}{PEPS}{projected entangled pair\hyp states}
\newacronym{ALS}{ALS}{alternating least squares}
\newacronym{BdG}{BdG}{Bogoljubov de-Gennes}
\newacronym{TFIM}{TFI}{transverse field Ising model}
\newacronym{PP}{PP}{projected purification}
\newacronym{BEC}{BEC}{Bose\hyp Einstein condensate}
\newacronym{JWT}{JWT}{Jordan\hyp Wigner transformation}
\newacronym{NISQ}{NISQ}{noisy intermediate scale quantum}
\newacronym{NN}{NN}{nearest\hyp neighbor}
\newacronym{NNN}{NNN}{next\hyp nearest\hyp neighbor}
\newacronym{SPDM}{SPDM}{single\hyp particle density matrix} 
\newacronym{HCB}{HCB}{hardcore bosons}
\newacronym{SF}{SF}{spinless fermions}
\newacronym{cQED}{cQED}{cirquit quantum electrodynamics}
\newacronym{HPC}{HPC}{high\hyp performance computing}
\newacronym{SSH}{SSH}{Su\hyp Schrieffer\hyp Heeger}
\newacronym{DSF}{DSF}{dynamical spin\hyp structure factor}
\newcommand{\acsaddress}{Department of Physics, Arnold Sommerfeld Center of Theoretical Physics, University of Munich, Theresienstrasse 37, 80333 Munich, Germany}
\newcommand{\mcqstaddress}{Munich Center for Quantum Science and Technology (MCQST), Schellingstrasse 4, 80799 M\"{u}nchen, Germany}

\newcommand{\nodagger}[0]{{\phantom{\dagger}}}

\Crefname{appendix}{Appendix}{Appendices}
\Crefname{equation}{Equation}{Equations}
\Crefname{figure}{Figure}{Figures}
\Crefname{section}{Section}{Sections}
\Crefname{tabular}{Tabular}{Tabulars}
\crefname{appendix}{App.}{Apps.}
\crefname{equation}{Eq.}{Eqs.}
\crefname{figure}{Fig.}{Figs.}
\crefname{section}{Sec.}{Secs.}
\crefname{tabular}{Tab.}{Tabs.}
\begin{document}
\title{Spectral decomposition and high\hyp accuracy Green's functions: \\ Overcoming the Nyquist\hyp Shannon limit via complex\hyp time Krylov expansion}
\author{S.~Paeckel}
\affiliation{\acsaddress}
\affiliation{\mcqstaddress}
\date{\today}
\begin{abstract}
  The accurate computation of low\hyp energy spectra of strongly correlated quantum many\hyp body systems, typically accessed via Green's\hyp functions, is a long\hyp standing problem posing enormous challenges to numerical methods.
  When the spectral decomposition is obtained from Fourier transforming a time series, the Nyquist\hyp Shannon theorem limits the frequency resolution $\Delta \omega$ according to the numerically accessible time domain size $T$ via $\Delta \omega = 2\pi/T$.
  In tensor network methods, increasing the domain size is exponentially hard due to the ubiquitous spread of correlations, limiting the frequency resolution and thereby restricting this ansatz class mostly to one\hyp dimensional systems with small quasi\hyp particle velocities.
  Here, we show how this limitation can be overcome by augmenting the time series with complex\hyp time Krylov states.
  At the example of the critical $S-1/2$ Heisenberg model and light bipolarons in the two\hyp dimensional~\acrlong{SSH} model, we demonstrate the enormous improvements in accuracy, which can be achieved using this method.
\end{abstract}
\maketitle
\paragraph*{Introduction}
The ability to compute the low\hyp energy spectra of strongly\hyp correlated quantum many\hyp body systems is of paramount importance in nearly every area of condensed matter physics.
In particular, the direct relation between spectral functions and Green's functions has proven extremely fruitful to foster our understanding of condensed matter systems.
This is based on the fact that spectral functions are directly accessible via various experimental probes such as ARPES, X-ray and neutron scattering, while Green's functions are mathematical objects probing the many\hyp body spectrum of the underlying Hamiltonians~\cite{benthien:256401,White2008,Pereira2009,Mourigal2013,Weber2015,Gohlke2017,Keselman2019,Grusdt2023,Kebric2024}.
Moreover, Green's functions are the essential building blocks of the most successfull methods to determine effective model Hamiltonians for the description of real materials~\cite{krishna-murthy,georges_RMP,bulla_DMFT,Kotliar2006,VCA1,SEF1,SEF2,VCA2,Werner2007,Reinhard2019,Bollmark2023,Lenz_2019,Karp2022,Bramberger2021,Bramberger2023,paeckel2023,Grundner2024a}.
The computation of Green’s functions spans complementary frameworks.
Diagrammatic and functional approaches provide controlled expansions in weak\hyp coupling regimes~\cite{Fradkin2013}.
Generalizations of density\hyp functional theory~\cite{Aryasetiawan1998,Onida2002} are highly successful but rely on the availability of proper exchange functionals, while non\hyp perturbative cluster embedding schemes are also applicable in the presence of strong correlations~\cite{Reuther2011,Metzner2012,Dupuis2021}.
Numerical methods provide an alternative route to deal with strongly correlated systems~\cite{Kuehner1999,Bulla2000,Aichhorn2003,Weichselbaum2007,Bulla2008}.
Among these~\glspl{TN}~\cite{White1992,Schollwoeck2010} are the de\hyp facto standard for one\hyp dimensional systems, while Monte\hyp Carlo~\cite{Prokofev1998,VanHoucke2010} is extraordinarily efficient in higher dimensions, yet suffers from the sign problem, especially for fermionic systems. 
Superficially, this methodical imbalance is attributed to the area law of entanglement, which, in two and higher dimensions, implies that already the representation of ground states exhibits an exponential scaling of the computational complexity~\cite{Eisert_RMP_arealaw,Eisert2015}.
However, the situation is even worse for~\gls{TN} based approaches because Green's functions are typically computed from time evolutions of excited states; and real\hyp time simulation are limited by the ubiquitous spread of correlations~\cite{LiebRobinson72}.
%
It is this additional growth of entanglement under time evolution, which drastically challenges their applicability to the computation of Green's functions in higher dimensions.
In this letter, we introduce a novel approach combining real and complex\hyp time evolutions to overcome this limitation.
At the example of the critical $S-1/2$ Heisenberg chain and light bipolaron formation in the two\hyp dimensional~\gls{SSH} model, we demonstrate the ability to significantly improve the frequency resolution of Green's functions, even for the case of two\hyp dimensional systems.
\paragraph*{Method}
We consider a general frequency\hyp dependent Green's function
\begin{equation}
  G_{AB}(\omega)
  =
  -\mathrm i \braket{\varphi_A | \left(H -E_0 - \omega - \mathrm i \eta\right)^{-1} | \varphi_B} \; , \label{eq:gf}
\end{equation}
where $\ket{\varphi_A},\ket{\varphi_B}$ are excited states, $\hat H$ denotes the Hamiltonian of the system under consideration with ground\hyp state energy $E_0$.
To regularize the poles of $G_{AB}$, a finite broadening $\eta > 0$ is introduced.
In~\gls{TN} simulations, $G_{AB}$ typically is computed by simulating the real\hyp time analogon and performing a Fourier transformation to frequency space
\begin{equation}
  G_{AB}(\omega)
  =
  \lim_{T\rightarrow \infty} \int_0^T dt \braket{\varphi_A | \hat U(t,\omega) | \varphi_B} \;, \label{eq:gf:t-domain}
\end{equation}
where we introduced $\hat U(t,\omega) = \mathrm e^{-\mathrm i (\hat H - E_0 - \omega - \mathrm i\eta)t}$ for convenience.
In practise, the time $T$ that can be reached is limited, such that the broadening $\eta$ has to be chosen to be of the order of $2\pi/T$ to smear out unphysical artefacts arising from the Fourier transformation on the finite time domain.
Let us recast~\cref{eq:gf:t-domain} by exploiting the group\hyp property of $\hat U(t,\omega)$ and split the integration domain into smaller intervals of size $T$, i.e., $\mathbb R^+_0 = [0,T) \cup [T,2T) \cup \cdots$.
We then introduce the operator $\hat S(T,\omega) = \sum_{p=0}^\infty \left[ \hat U(T, \omega) \right]^p$, which boosts the evolution on the finite interval $[0,T)$ to all intervals $[pT,(p+1)T)$ such that
\begin{align}
  G_{AB}(\omega)
  &=
  \int_0^T dt \braket{\varphi_A | \hat U(t,\omega) \hat S(T,\omega) |\varphi_B}\;. \label{eq:gf:geometric-sum}
\end{align}
The geometric series in $\hat S(T,\omega)$ always converges as long as $\eta T > 0$, while a truncation at zeroth order $p=0$ yields $\hat S = 1$, and one recovers the usual finite\hyp time integration over the domain $[0,T]$.
In that case, the Nyquist\hyp Shannon theorem~\cite{Nyquist1928,Shannon1949} dictates that only an approximation to the Green's function $\tilde G^T_{AB}(\omega=\omega_k)$ sampled at frequencies $\omega_k = \frac{2\pi}{T}k \equiv \Delta \omega\cdot k$ ($k\in\mathbb Z$) is obtained.
It is now convenient to define $\hat K = \hat S - 1$.
This allows to decompose the full Green's function into the $p=0$ part $\tilde G^T_{AB}$, and a correction $K^T_{AB}$
\begin{align}
  G_{AB} &= \tilde G^T_{AB} + K^T_{AB} \; , \label{eq:gf:decomposed} \\
  K^T_{AB} &= \int_0^T dt \braket{\varphi_A | \hat U(t,\omega) \hat K(T,\omega) |\varphi_B} \; .
\end{align}
Harnessing additional information about the spectral properties encoded in $K^T_{AB}(\omega)$, also faint signals can be recovered, contrasting this approach to compressed sensing schemes~\cite{Tao2005,Tao2006}
The task is now to determine $K^T_{AB}$ and we show in the following how this contribution can be approximated efficiently using~\glspl{TN}.
The key observation is that $K^T_{AB}$ accounts for the low\hyp frequency part of $G_{AB}$.
In fact, setting $\hat S \equiv 1$ essentially introduces a low\hyp pass filter on the time domain such that the Green's function in frequency space is fully determined from equally spaced samples $\tilde G^T_{AB}(\omega_k)$~\cite{Nyquist1928,Shannon1949}.
Therefore, information about the dependency of $G_{AB}$ on smaller frequencies $\omega\in[0,\Delta\omega]$ must be encoded in $K^T_{AB}$.
Moreover, the Green's function typically exhibits a finite spectral width, i.e., it is non\hyp zero only in a finite frequency interval because the states $\ket{\varphi_{\mu}}$ with $\mu=A,B$ are mostly obtained from local excitations above the ground state of $\hat H$.
Thus, $\hat K(T,\omega)$ is ideally evaluated in a low\hyp energy subspace.
In general, low\hyp energy subspaces are obtained by constructing Krylov spaces via successively applying $\hat H$ to proper initial states $\ket{\varphi_{\mu}}$.
However, the low\hyp energy resolution in such a Krylov space converges slowly for critical systems or those with small gaps, hence limiting the applicability significantly~\cite{lanczos,lanczos_book,Freund2000}.
We therefore pursue another approach building a Krylov space from time\hyp evolving the initial states along a complex contour of segments $\tau = (1-\mathrm i\tan(\alpha)) \delta t$.
The time\hyp evolution operator for each segment can be decomposed
\begin{equation}
  \hat U_\tau = \mathrm e^{-\hat H \delta t\tan(\alpha)} \mathrm e^{-\mathrm i \hat H \delta t} \; , \label{eq:complex-tevo}
\end{equation}
for a complex angle $\alpha \in [0,\pi/2)$.
Here, the role of the exponential damping $\mathrm e^{-\hat H \delta t\tan(\alpha)}$ is to suppress higher energy contributions such that the Krylov space
\begin{equation}
  \mathcal H^{D\tau}_\mu = \operatorname{span}\left\{\frac{\ket{\varphi_\mu}}{\lVert \ket{\varphi_\mu}\rVert}, \frac{\hat U_\tau\ket{\varphi_\mu}}{\lVert \hat U_\tau\ket{\varphi_\mu} \rVert}, \cdots, \frac{\hat U^{D-1}_\tau \ket{\varphi_\mu}}{\lVert \hat U^{D-1}_\tau \ket{\varphi_\mu} \rVert} \right\} \; ,
\end{equation}
is spanned by normalised states that exhibit a large overlap with the low\hyp energy eigenstates of $\hat H$.
The complex contour has immediate effects on the efficiency of~\gls{TN} methods.
The continuous suppresion of high\hyp energy states yields a significant reduction of bond dimension upon time evolving the initial states.
This has been exploited only recently to improve the efficiency of time evolutions, yet the complex time was promoted to a complex frequency by a direct Fourier transform, requiring further postprocessing or evaluations of expensive auxiliary quantities to obtain real\hyp frequency data~\cite{Grundner2024b,Cao2024}.
In contrast, in our approach we use the complex\hyp time Krylov space $\mathcal H^{D\tau}_\mu$ to expand the Hamiltonian $\hat H$ directly.
Approximating $\hat K(T,\omega)$ on the real frequency axis then allows to evaluate the time integration in the $T\rightarrow \infty$ limit faciliating arbitrary frequency resolution.
Given a collection of states $\left\{\ket{\varphi^0_B}, \ket{\varphi^1_B}, \ldots\right\}$ generated via complex\hyp time evolution of the initial state $\ket{\varphi^n_B} = \hat U_{n\tau}\ket{\varphi_B}/\lVert \hat U_{n\tau}\ket{\varphi_B} \rVert$, the following steps are required to evaluate $K^T_{AB}$:
\begin{enumerate}
  \item
    Construct the Gram matrix $\mathbf M$ with elements $M_{nm} = \braket{\varphi^n_B|\varphi^m_B}$ (and denote by $\vec{M}_m$ the $m$th column vector of $\mathbf M$).
  \item
    Diagonalize $\mathbf M = \mathbf{U D U}^\dagger$ and compute $\mathbf X = \mathbf U \mathbf D^{-\nicefrac{1}{2}}$ to get a transformation into an orthonormal basis $\ket{\psi^i_B} = \sum_{n=1}^r X_{ni} \ket{\varphi^n_B}$. If there are eigenvalues $D_{ii} \leq 0$, discard them to ensure the basis has full rank $r \leq D$.
  \item
    Compute $\tilde H^\mathrm{eff}_{nm} = \braket{\varphi^n_B|\hat H|\varphi^m_B}$ and transform the matrix representation of the effective Hamiltonian $\tilde{\mathbf H}^\mathrm{eff}$ into the orthonormal basis $H^\mathrm{eff}_{ij} = \braket{\psi^i_B|\hat H|\psi^j_B} = (\mathbf X^* \tilde{\mathbf H}^\mathrm{eff} \mathbf X^t)_{ij}$.
  \item
    Compute $Y_{AB,n}(t)=\braket{\varphi_A(-t)|\varphi^n_B}$ where $\ket{\varphi_A(t)}$ are the real\hyp time evolutions of $\ket{\varphi^A}$.
\end{enumerate}
Using the eigendecomposition $\mathbf H^\mathrm{eff}=\mathbf Q \mathbf E^\mathrm{eff} \mathbf Q^\dagger$, the final expression for $K^T_{AB}(\omega)$ is given by
\onecolumngrid
\begin{align}
  K^T_{AB}(\omega)
  &=
  \int_0^T dt \vec{Y}_{AB}(t) \mathrm e^{\mathrm i(E_0+\omega+\mathrm i\eta)t} \mathbf X \mathbf Q \frac{\mathrm e^{-\mathrm i(\mathbf E^\mathrm{eff} - E_0 - \omega - \mathrm i\eta)T}}{1-\mathrm e^{-\mathrm i(\mathbf E^\mathrm{eff} - E_0 - \omega - \mathrm i\eta)T}} \mathbf Q^\dagger \mathbf X^\dagger \vec{M}_0 \; . \label{eq:K}
\end{align}
\twocolumngrid
The costs to evaluate~\cref{eq:K} are dominated by the scalar products to obtain $M_{nm}$, $\tilde H^\mathrm{eff}_{nm}$ and $Y_{AB,n}(t)$.
Here, the most expensive part is the computation of the $Y_{AB,n}(t)$'s, which have to be done for all real and complex\hyp time evolved states $\ket{\varphi_A(t)}$ and $\ket{\varphi^m_B}$, respectively.
Luckily, these are all independent on each other, can be parallelized trivially, and the actual~\gls{TN} operations to evaluate overlaps are subleading compared to the time evolutions~\cite{Paeckel2019}.
The convergence of the Green's function augmented by the complex\hyp time Krylov space correction~\cref{eq:K} can be analyzed in terms of Arnoldi iterations.
A rigorous upper bound for the approximation error is derived in~\footnote{See Supplemental Material at [URL will be inserted by publisher] for a detailed error analysis, which includes Refs. \cite{lubich_errors,lubich_timeevolve,Stoudenmire:2013eq,Shen2023}} yielding the practically relevant noise threshold
\begin{equation}
  R \lesssim \frac{2}{\delta t \eta D} h_{r,r-1} \;, \label{eq:noise-threshold}
\end{equation}
where $h_{r,r-1}$ is the Arnoldi residual, which can be computed directly from $\mathbf H_\mathrm{eff}$, and $D$ denotes the spectral width of $G_{AB}$~\footnote{Bringing $\mathbf H_\mathrm{eff}$ into upper Hessenberg form directly allows to read of the matrix element $h_{r,r-1}$}.
Crucially, $h_{r,r-1}$ is exponentially suppressed $\sim \mathrm e^{r(1+\ln(\eta\tilde\lambda))}$ as long as $\ln(\delta t\tilde\lambda) < -1$, where $\tilde\lambda$ is the spectral width of $\ket{\varphi_B}$.
Thus, for large rank $r$ of the Gram matrix as well as small time steps $\delta t$, an exponential convergence towards the exact Green's function can be expected.
Increasing the complex angle $\alpha$ is also beneficial, yet care must be taken: $\alpha>0.1$ should be avoided to prevent numerical instabilities, and to ensure linear independence for the Krylov states, i.e., sufficiently large $r$.
\paragraph*{$S-1/2$ Heisenberg chain}
\begin{figure}
  \centering
  \subfloat[\label{fig:heisenberg:spec}]{
    \includegraphics[width=0.49\textwidth,trim={0 0 0 3.5em},clip]{\figurepath/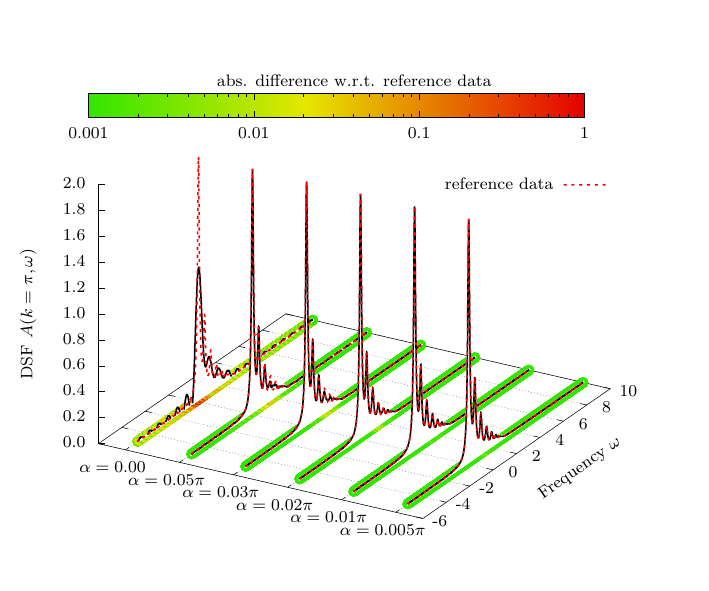}
  }\vspace*{-2.5em}
  \\
  \subfloat[\label{fig:heisenberg:convergence}]{
    \includegraphics[width=0.49\textwidth]{\figurepath/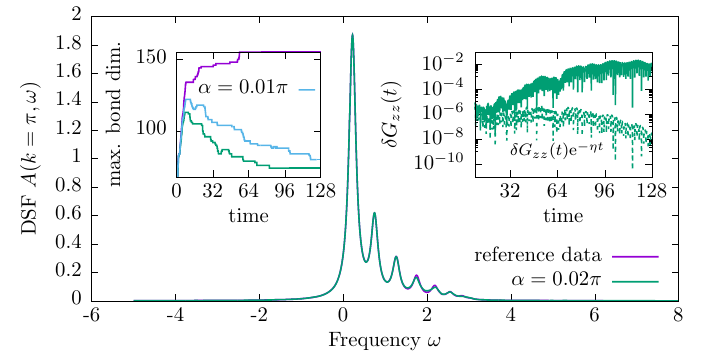}
  }
  \caption{
    \label{fig:heisenberg}
    \subfigref{fig:heisenberg:spec}: Comparison between~\acrshortpl{DSF} of the $S-1/2$ Heisenberg model for different complex angles $\alpha$ (black) and a max. complex simulation time $\mathrm {Re}(T_\mathrm{cplx})=128$ with exact reference data (red).
    Real\hyp time evolution was done until $T=8$ and $\alpha=0$ shows the uncorrected~\acrshort{DSF}.
    The absolute difference between the complex time and the exact reference data is indicated by the color\hyp coded baselines.
    \subfigref{fig:heisenberg:convergence}~\acrshort{DSF} for $\alpha=0.02\pi$ to illustrate the precision of the corrected Green's function.
    Left inset displays the bond dimensions of the complex\hyp time evolution (green, blue), compared to the real case (purple).
    Right inset shows the error of the local time\hyp dependent Green's function when propagating the time evolution beyond $T=8$ using the boost operator $\hat S(T,\omega)$ (solid line) and after incorporating the damping (dashed line) generated by the finite broadening $\eta=0.1$.
  }
\end{figure}
As a first testcase, we consider Green's functions of the isotropic $S-1/2$ Heisenberg chain
\begin{equation}
  \hat H =  \sum_j \hat{\vec S}_j \hat{\vec S}_{j+1} \label{eq:model:heisenberg} \;.
\end{equation}
The model is critical, i.e., there are gapless spinon excitations challenging most Krylov expansion schemes.
To faithfully study the effect of incorporating the contributions $K^T_{AB}(\omega)$, we set the system size to $L=16$ spins, which can be solved numerically exactly.
We computed the~\gls{DSF} $A(k,\omega) = -\frac{1}{\pi}\operatorname{Im} G_{zz}(k,\omega)$, where
\begin{equation}
  G_{zz}(k,\omega) = \braket{k|\left[\hat H-E_0-\omega-\mathrm i\eta\right]^{-1}|k} \;,
\end{equation}
and $\ket{k} = \frac{1}{\sqrt L}\sum_j \mathrm e^{-\mathrm i k r_j} \hat S^z_j \ket{E_0}$ is a longitudinal momentum space excitation of the ground state $\ket{E_0}$.
Using a~\gls{MPS} representation~\cite{White1992,White1993,Schollwoeck2005,Schollwoeck201196}, we performed real\hyp time simulations~\cite{Paeckel2019} of the excited states $\ket{k}$ until a time $T=8$, corresponding to the time frame in which a spinon excitation traverses the system.
Since spinons can scatter off each other, larger simulation times yield a significant increase of correlations, which in more complex systems would cause an exponential growth in the~\gls{MPS} bond dimension at a fixed truncation error~\cite{Paeckel2019} (here we allowed $\delta=10^{-10}$ as the maximal discarded weight per site).
It is exactly this breakdown of~\gls{TN} based methods, which we alleviate by limiting the real\hyp time evolution and instead involve~\cref{eq:K}.
To evaluate~\cref{eq:K}, we performed complex\hyp time evolutions of the initial states for several complex angles $\alpha$ until a maximum time $\mathrm {Re}(T_\mathrm{cplx})=128$.
For each value of $\alpha$, we then constructed the Krylov subspace from maximally $D=129$ time evolved states (including the initial state $\ket{k}$) using an equally spaced grid along the complex\hyp time contour with $\delta t=1$.
The~\gls{DSF} is then obtained from~\cref{eq:K} and we chose a broadening $\eta=0.1$.
In~\cref{fig:heisenberg:spec} we compare against the numerically exact~\gls{DSF} (red).
The curve with $\alpha=0$ shows results using the finite\hyp time integration, only.
Note the artificial oscillations as well as the stark deviations in the peak height at $\omega=0$.
This is also indicated by the color\hyp coded baseline displaying the absolute difference to the exact~\gls{DSF}.
Noteworthy, as soon as $K^T_{AB}(\omega)$ is added (finite $\alpha$), the oscillations dissapear, which is an immediate consequence of the fact that incorporating~\cref{eq:gf:geometric-sum} we work in the limit $T\rightarrow\infty$.
Furthermore, even for the largest complex angle $\alpha=0.05\pi$, the low\hyp frequency behavior of the~\gls{DSF} is already well captured and deviations are visible only at larger frequencies $\omega > 1$.
These errors vanish when decreasing $\alpha$, until for $\alpha=0.02\pi$ we find excellent agreement, which is demonstrated in~\cref{fig:heisenberg:convergence}, too.
In the left inset the computational benefits are shown by comparing the max. bond dimension during the complex\hyp time evolution to that occuring in a conventional real\hyp time evolution.
We also tested the quality of the approximation of $\hat K$ by explicitly forecasting the dynamics of the local excitations.
In the right inset we plot the absolute error of the dynamics of the time\hyp dependent Green's function $G_{zz}(t)=\braket{E_0|\hat S^z_8(t)\hat S^z_8|E_0}$ starting the prediction at a time $T=8$.
Note that even at very long times $T>100$ the errors are small $\lesssim 10^{-2}$ and in particular revivals caused from finite size effects, i.e., reflections of spinon excitations at the boundaries, are treated correctly such that incorporating the damping (dashed line) yields extremely high precision.
\paragraph*{Two\hyp dimensional SSH\hyp model}
\begin{figure}
  \centering
  \subfloat[\label{fig:ssh:spec:alpha-small:s0}]{
    \includegraphics[width=0.49\textwidth,trim={0.5em 0.5em 1.5em 3.5em},clip]{\figurepath/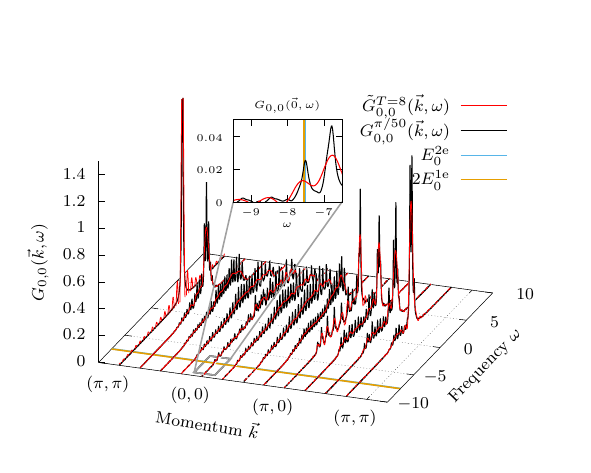}
  }
  \\
  \subfloat[\label{fig:ssh:spec:alpha-large:bipolaron}]{
    \includegraphics[width=0.48\textwidth]{\figurepath/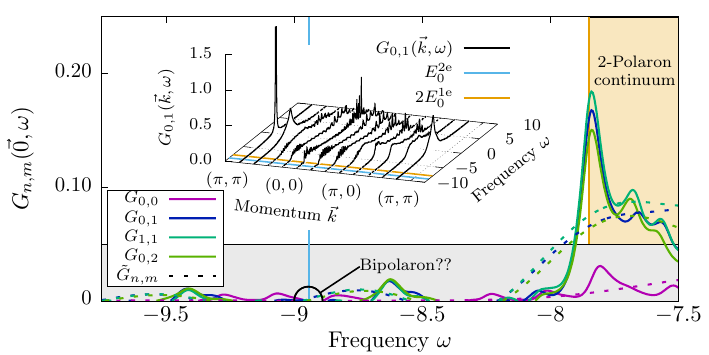}
  }
  \caption{
    \label{fig:ssh}
    \subfigref{fig:ssh:spec:alpha-small:s0}: Comparison between uncorrected $S0$\hyp bipolaron Green's function $\tilde G^{T=8}_{0,0}$ and complex\hyp time Krylov space augmented $G^{\pi/50}_{0,0}$ with $\alpha=\pi/50$ on a $8\times 8$ cluster for $\lambda=0.025$, following a high\hyp symmetry path through the Brillouin zone.
    Note the distinct peak structure indicating dressed two\hyp electron states, which becomes visible only when taking into account $K^T_{AB}(\omega)$.
    Inset: The corrected $\vec k=\vec 0$ Green's function exhibits a distinguished peak right at the ground\hyp state energy $\omega=E^\mathrm{2e}_0$ computed via~\acrshort{DMRG}, which is completely washed out in $\tilde G^{T=8}_{0,0}$.
    \subfigref{fig:ssh:spec:alpha-large:bipolaron}: Corrected Green's functions $G^{\pi/33}_{n,m}(\vec 0, \omega)$ for different electron displacements $\vec a_{n,m}=n\vec e_x + m\vec e_y$ in the strong coupling regime $\lambda=0.2$ at small frequencies.
    Vertical blue curve indicates the two\hyp electron ground\hyp state energy $E^\mathrm{2e}_0$, which is well separated from the two\hyp polaron continuum at frequencies $\omega>2E^\mathrm{1e}_0$.
    Above the noise threshold $R\approx 0.05$ (gray shaded area) no signal for a bipolaronic bound state can be found.
    Inset shows $G^{\pi/33}_{0,1}(\vec 0,\omega)$ along a high\hyp symmetry path through the Brillouin zone.
  }
\end{figure}
We now turn to a more challenging testcase to demonstrate the power of our method.
We study isolated two\hyp electron bound states (bipolarons) in the Peierls~\gls{SSH} model~\cite{Baris1972a,Baris1972b,SSH_1979} on a two\hyp dimensional square lattice, in which electrons couple non\hyp locally to optical phonons.
The Hamiltonian is given by
\begin{align}
  \hat H
  &=
  -\sum_{\langle \vec r, \vec r^\prime\rangle}\sum_{\sigma} \left(\hat c^\dagger_{\vec r,\sigma}\hat c^\nodagger_{\vec r^\prime,\sigma} + \mathrm{h.c.} \right) \left(t - g\hat V^\nodagger_{\vec r, \vec r^\prime} \right) \notag \\
  &\phantom{=}\quad
  +U\sum_{\vec r} \hat n^\nodagger_{\vec r,\uparrow} \hat n^\nodagger_{\vec r,\downarrow} + \Omega\sum_{\vec r} \hat a^\dagger_{\vec r} \hat a^\nodagger_{\vec r} \; , \label{eq:ssh-model}
\end{align}
where $\hat c^{(\dagger)}_{\vec r,\sigma}$ are the spin\hyp $\sigma$ electronic ladder operators, $\hat a^{(\dagger)}_{\vec r}$ denote the annihilation (creation) operators of optical phonons with frequency $\Omega$ and the sum $\sum_{\langle \vec r, \vec r^\prime\rangle}$ is over nearest neighboring lattice sites $\vec r, \vec r^\prime$.
The interaction is given by $\hat V_{\vec r, \vec r^\prime} = \sqrt{2 M \Omega} \left( \hat X^\nodagger_{\vec r} - \hat X^\nodagger_{\vec r^\prime} \right)$ with the displacements $\hat X^\nodagger_{\vec r}=\frac{1}{\sqrt{2M\Omega}}(\hat a^\dagger_{\vec r} + a^\nodagger_{\vec r})$, and we chose the oscillator mass $M = 1$ as well as $t=1$ as unit of energy.
Recently, this model and variations of it have gained attention due to the predicted existance of light bipolarons, which could give rise to a phonon\hyp mediated mechanism for high\hyp$T_c$ superconductivity~\cite{Sous2018,Zhang2023,grundner2023,Kim2024}.
Unfortunately, investigating the spectral properties of~\cref{eq:ssh-model} poses significant challenges to state\hyp of\hyp the\hyp art Monte\hyp Carlo methods.
\Gls{TN} investigations have been far out of reach, even the case of isolated bipolarons in one dimension is extremely challenging~\cite{grundner2023}.
Here, we study the practically relevant case of small phonon frequencies (adiabatic limit) $\Omega/t=0.2$ on finite lattices of size $8\times 8$.
We chose a finite Hubbard repulsion $U/t=2$ and simulated weak and strong interactions $\lambda = \frac{g^2}{2\Omega t} = 0.025,0.2$.
In order to investigate bipolaronic quasi\hyp particles, we computed the two\hyp electron Green's functions
\begin{align}
  G_{n,m}(\vec k,\omega)
  &=
  \braket{\vec k,n,m|\left[\hat H-\omega-\mathrm i\eta\right]^{-1}|\vec k,n,m} \;, \label{eq:ssh:gf}
\end{align}
where $\ket{\vec k,n,m} = \frac{1}{\sqrt L} \sum_{\vec r} \mathrm e^{-\mathrm i \vec k (2\vec r+\vec a_{n,m})}\hat c^\dagger_{\vec r,\uparrow} \hat c^\dagger_{\vec r + \vec a_{n,m},\downarrow} \lvert \varnothing \rangle$ and $\vec a_{n,m} = n\vec e_x + m\vec e_y$.
The excitations $\ket{\vec k,n,m}$ correspond to two\hyp electron states with total momentum $\vec k$, being spatially separated by $\vec a_{n,m}$, and the Green's functions~\cref{eq:ssh:gf} probe the contributions of these excitations to the eigenstates of~\cref{eq:ssh-model}~\cite{Berciu2011}.
We employ the~\gls{PP}~\cite{Koehler2021,Stolpp2021} to account for the phononic Hilbert space and a single\hyp site variant of the~\gls{TDVP} to perform the time evolutions~\cite{PhysRevLett.107.070601,PhysRevB.94.165116,Yang_2020prb,grundner2023}, using a max. discarded weight $\delta = 10^{-10}$ and a max.~\gls{MPS} bond dimension $m=4096$.
Real\hyp time evolutions were done until a maximum time $T=8$, longer times are practically inaccesible due to the exponential growth of the bond dimension~\footnote{In fact, the maximum time $T=8$ could be reached only for few initial states, due to the rapid growth of memory requirements.}.
For the complex\hyp time evolution, we set the maximum time to $\mathrm {Re}(T_\mathrm{cplx})=64$ using time step sizes $\delta t=0.1$, and varied the complex angle between $\alpha=0.02\pi$ and $\alpha=0.05\pi$.
In~\cref{fig:ssh:spec:alpha-small:s0} we present results evaluating the imaginary part of~\cref{eq:ssh:gf} for the case of $S0$ bipolarons ($n=m=0$) with $\lambda=0.025$ following a high\hyp symmetry path through the Brillouin zone (we use a broadening $\eta=0.1$).
We compare the uncorrected Green's function $\tilde G^{T=8}_{0,0}(\vec k,\omega)$ (red) to the complex\hyp time corrected Green's function $G^{\pi/50}_{0,0}(\vec k, \omega)$ (black) taking into account~\cref{eq:K} and a complex angle $\alpha=0.02\pi$.
Note the peculiar peak structure, which is completely absent in $\tilde G^{T=8}_{0,0}(\vec k,\omega)$.
The peaks are spaced $\approx k\Omega$ ($k \in \mathbb N$) and thus explicitly demonstrate the formation of polaronic bound states.
We verified that the peaks are no artefacts by performing~\gls{DMRG} simulations.
The one\hyp/two\hyp electron ground\hyp state energy $E^\mathrm{1e}_0/E^\mathrm{2e}_0$ is indicated by the yellow/blue line and in the inset we show a zoom on the Green's functions in the vincinity of $\omega=E^\mathrm{2e}_0$, for $\vec k = \vec 0$~\footnote{Due to the finite cluster size there are strong finite size effects. We checked this by computing the ground state of the non\hyp interacting model yielding a ground\hyp state energy, which is always larger than the one obtained when incorporating finite $\lambda$}.
Strikingly, for the two\hyp electron case $G^{\pi/50}_{0,0}$ shows a pronounced peak centered exactly at $\omega=E^\mathrm{2e}_0$ while $\tilde G^{T=8}_{0,0}$ exhibits only an unspecific oscillation.
Moreover, the strong peak at $\vec k=(\pi,\pi)$ is generated by another exact eigenstate of~\cref{eq:ssh-model}.
Note how taking into account the complex\hyp time correction mitigates artefacts in the real\hyp time evolved Green's function and produces a smooth and sharp signal.
This increase in accuracy allows to deduce that, up to finite size effects and upon evaluating the binding energy as $E^\mathrm{2e}_0 - 2E^\mathrm{1e}_0$, for weak coupling $\lambda=0.025$ there is no bipolaronic ground state.
Instead, we only observe unbound two\hyp polaron states with $w > 2E^\mathrm{1e}_0$, i.e., a two\hyp polaron continuum.
The situation changes in the strong coupling regime $\lambda=0.2$ (i.e., $\lambda/\Omega = 1)$.
In~\cref{fig:ssh:spec:alpha-large:bipolaron} we compare the imaginary part of the complex\hyp time Krylov space agumented Green's functions~\cref{eq:ssh:gf} at $\vec k = \vec 0$ varying the spatial separation $\vec a_{n,m}$ between $\vec a_{n,m}=(0,0)$ and $\vec a_{n,m}=(0,2)$, using a broadening $\eta=0.1$ and $\mathrm {Re}(T_\mathrm{cplx})=32$.
In that case, the two\hyp electron ground\hyp state energy $E^\mathrm{2e}_0$ (vertical blue line) obtained via~\gls{DMRG} is well separated from the two\hyp polaron continuum $2E^\mathrm{1e}_0$ (yellow vertical line).
Notably, we do not observe a feature at $E^\mathrm{2e}_0$ for any of the considered electron displacements $\vec a_{n,m}$ above the noise threshold $R\approx 0.05$, which indicates that the trial states $\ket{\vec k,n,m}$ do not contribute significantly to the two\hyp electron ground state.
A possible explanation for this suprising finding is that the many\hyp body ground state, if representing a bipolaron, are much stronger correlated to their surrounding phonon clouds then previously anticipated~\cite{Berciu2011}.
We carefully checked in~\footnote{See Supplemental Material at [URL will be inserted by publisher] for a detailed convergence analysis} that the oscillatory features visible below the noise threshold, for instance near $\omega = -8.6$, are in fact replicas, i.e., remaining artefacts not accounted for by the correction term.
\paragraph*{Discussion/Outlook}
We introduced a reformulation of the Fourier transformation of a general Green's function from time to frequency domain~\cref{eq:gf:geometric-sum}, decomposing the integrand into a real\hyp time evolution on a finite time domain and a correction $\hat K$.
Constructing a complex\hyp time Krylov space, we demonstrated that $\hat K$ can be approximated efficiently using~\gls{TN} methods, allowing us to overcome the Nyquist\hyp Shannon sampling limit in frequency space.
The resulting Green's functions exhibit a significantly higher accuracy, even in extremely challenging two\hyp dimensional systems, opening the path to a precise spectral analysis of models that have been way beyond the reach of~\gls{TN} methods, so far.
However, we expect that our method can be further improved in various ways.
For instance, in this letter we only use a fixed complex angle.
Incorporating the entanglement growth along the contour could yield more efficient evolution schemes to further increase the complex\hyp time Krylov subspace dimension.
Moreover, our error analysis suggests that direct sums of Krylov subspaces constructed from states evolved along different complex contours could drastically improve the convergence rate.
\paragraph*{Acknowledgements}
We thank Philipp Westhoff and John Sous for insightful discussions about the~\gls{SSH} model, and Ulrich Schollw\"ock, Thomas K\"ohler and Martin Grundner for carefully reading the manuscript.
The author acknowledges support by the Deutsche Forschungsgemeinschaft (DFG, German Research Foundation) under Germany’s Excellence Strategy-426 EXC-2111-390814868.
This work was supported by Grant No. INST 86/1885-1 FUGG of the German Research Foundation (DFG).
The data in this work was produced using \syten~\cite{hubig:_syten_toolk,hubig17:_symmet_protec_tensor_networ} and \symmps~\cite{symmps}.
\bibliography{literature}
\end{document}


\title{Appendix: Spectral decomposition and high\hyp accuracy Green's functions: \\ Overcoming the Nyquist\hyp Shannon limit via complex\hyp time Krylov expansion}
\author{S.~Paeckel}
\affiliation{\acsaddress}
\affiliation{\mcqstaddress}
\date{\today}
\maketitle
\setcounter{equation}{0}%
\setcounter{figure}{0}%
\setcounter{table}{0}%
\setcounter{page}{1}%
\makeatletter%
\renewcommand{\theequation}{A\arabic{equation}}%
\renewcommand{\thefigure}{A\arabic{figure}}%
\appendix
\numberwithin{equation}{section}
\section{Detailed derivation of the complex\hyp time Krylov subspace approximation of $G_{AB}(\omega+\mathrm i\eta)$\label{app:sec:derivation}}
We consider a general frequency\hyp dependent Green's function
\begin{equation}
  G_{AB}(\omega)
  =
  -\mathrm i \braket{\varphi_A | \left(H -E_0 - \omega - \mathrm i \eta\right)^{-1} | \varphi_B} \; , \label{app:eq:gf}
\end{equation}
where $\ket{\varphi_A},\ket{\varphi_B}$ are excited states, $\hat H$ denotes the Hamiltonian of the system under consideration, and $\eta > 0$ is a finite broadening.
The eigenstates $\ket{E_n}$ of $\hat H$ are labeled by the eigenvalues $E_0 \leq E_1 \leq E_2 \leq \cdots$ with the ground\hyp state energy $E_0$.
Using the definition $\hat U(t,\omega) = \mathrm e^{-\mathrm i (\hat H - E_0 - \omega)t}$,~\cref{app:eq:gf} can be rewritten as
\begin{equation}
  G_{AB}(\omega)
  =
  \int_0^\infty \mathrm dt\, \braket{\varphi_A|\hat U(t,\omega+\mathrm i\eta)|\varphi_B}
  =
  \int_0^T \mathrm dt\, \braket{\varphi_A|\hat U(t,\omega+\mathrm i\eta) \sum_{p=0}^\infty \left[\hat U(T,\omega + \mathrm i\eta)\right]^p |\varphi_B} \;, \label{app:eq:gf-inf}
\end{equation}
for some $T>0$.
We refer to $\hat S(T,\omega+\mathrm i\eta) = \sum_{p=0}^\infty \left[\hat U(T,\omega + \mathrm i\eta)\right]^p = \left[1-\hat U(T,\omega + \mathrm i\eta)\right]^{-1}$ as boost operator, translating the time evolution by multiples of the integration domain size $T$.
Let us now consider the complex\hyp time Krylov subspace $\mathcal K^{D\tau}_\mu$ generated by successively evolving the state $\ket{\varphi_\mu}$ with step size $\tau$ along a complex contour
\begin{equation}
  \mathcal K^{D\tau}_\mu
  =
  \operatorname{span}\left\{\frac{\ket{\varphi_\mu}}{\lVert \ket{\varphi_\mu} \rVert}, \frac{\hat U_\tau\ket{\varphi_\mu}}{\lVert \hat U_\tau\ket{\varphi_\mu}\rVert}, \cdots, \frac{\hat U^{D-1}_\tau \ket{\varphi_\mu}}{\lVert \hat U^{D-1}_\tau \ket{\varphi_\mu}\rVert} \right\}
  \equiv
  \operatorname{span}\left\{\ket{\varphi^0_\mu}, \ket{\varphi^1_\mu}, \cdots,\ket{\varphi^{D-1}_\mu} \right\} \; ,
\end{equation}
where $\hat U_\tau = \mathrm e^{-\mathrm i \hat H \tau}$ and in the following we choose $\tau = (1-\mathrm i \tan \alpha) \delta t$ with $\alpha \in [0,\pi/2)$ denoting the angle between the complex contour and the real\hyp time axis, and $\delta t > 0$.
For later convenience, we introduce the projector into the Krylov subspace $\hat P^\mathcal{D\tau}_B$.
We want to approximate the action of the boost operator in the subspace $\mathcal K^{D\tau}_B$.
For that purpose, we expand the Hamiltonian in terms of an orthonormal basis $\left\{\ket{\psi^i_B}\right\}$ of $\mathcal K^{D\tau}_B$, where $i = 1,\ldots r$ and $r = \operatorname{dim} \mathcal K^{D\tau}_B \leq D$ is the dimension of the Krylov subspace.
Such an orthonormal basis can be obtained from the Gram matrix $\mathbf M$ with elements $M_{ij} = \braket{\varphi^i_B|\varphi^j_B}$, via
\begin{equation}
  \ket{\psi^i_B} = \sum_{n=0}^{D-1} X_{ni} \ket{\varphi^n_B}\; , \quad \text{where} \quad
  \mathbf M = \mathbf U \mathbf D \mathbf U^\dagger \quad \text{and} \quad \mathbf X = \mathbf U \mathbf D^{-\nicefrac{1}{2}} \mathbf U^\dagger \; .
\end{equation}
Here, the matrix $\mathbf X$, whose columns contain the coefficients of the basis transformation, is computed from the square root of the pseudo inverse of the Gram matrix.
Expanding the Hamiltonian in this basis, i.e., $H^\mathrm{eff}_{ij} = \braket{\psi^i_B|\hat H|\psi^j_B}$, we can approximate the boost operator in~\cref{app:eq:gf-inf}
\begin{align}
  G_{AB}(\omega)
  &\approx
  \int_0^T \mathrm dt\, \bra{\varphi_A}\hat U(t,\omega+\mathrm i\eta) \sum_{ij=1}^r \ket{\psi^i_B}\braket{\psi^i_B|\hat S(T,\omega+\mathrm i\eta)|\psi^j_B}\braket{\psi^j_B|\varphi_B} \\
  &=
  \int_0^T \mathrm dt\, \bra{\varphi_A}\hat U(t,\omega+\mathrm i\eta) \sum_{ij=1}^r \ket{\psi^i_B}\braket{\psi^i_B|\left[1-\hat U(T,\omega+\mathrm i\eta)\right]^{-1}|\psi^j_B}\braket{\psi^j_B|\varphi_B} \\
  &\approx
  \int_0^T \mathrm dt\, \bra{\varphi_A}\hat U(t,\omega+\mathrm i\eta) \left[\sum_{ij=1}^r \ket{\psi^i_B}
  \left(1 - \mathrm e^{-\mathrm i (H^\mathrm{eff}_{ij} - E_0 - \omega - \mathrm i \eta)T}\right)
  \bra{\psi^j_B}\right]^{-1} \ket{\varphi_B} \; .
  \label{app:eq:boost-operator:approx}
\end{align}
In the last line, we used the approximation of operator exponentials in the Krylov subspace $\mathcal K^{D\tau}_B$
\begin{equation}
  \hat P^\mathcal{D\tau}_B
  \mathrm e^{-\mathrm i (\hat H-E_0-\omega-\mathrm i \eta)pT}
  \hat P^\mathcal{D\tau}_B
  \approx
  \mathrm e^{-\mathrm i (\hat P^\mathcal{D\tau}_B\hat H \hat P^\mathcal{D\tau}_B - E_0-\omega-\mathrm i \eta)pT} \; . \label{app:eq:exp-approx}
\end{equation}
For the case of a Krylov subspace generated from $\hat H$, error bounds for this approximation have been derived~\cite{lubich_errors,lubich_timeevolve}.
Here, however, the Krylov subspace is constructed from $\hat U_\tau$ and to the best of our knowledge, error estimations for~\cref{app:eq:exp-approx} have not been studied so far.
For that reason,~\cref{app:sec:error-analysis} is devoted to the analysis of the used approximations, and we derive rigorous error bounds for the approximation of the Green's function.
The numerical evaluation of the operator inverse in~\cref{app:eq:boost-operator:approx} constitutes the main novelty of the algorithm and its efficient computation is crucial for the success of the presented method.
This means in particular that an explicit construction of the orthonormal basis states $\ket{\psi^i_B}$ should be avoided.
Defining the overlap matrix $Y_{AB,n}(t,\omega+\mathrm i\eta) = \braket{\varphi_A|\hat U(t,\omega+\mathrm i\eta|\varphi^n_B}$ as well as the eigenvalue decomposition of the effective Hamiltonian $\mathbf H^\mathrm{eff} = \mathbf Q \mathbf E^\mathrm{eff} \mathbf Q^\dagger$, we can rewrite~\cref{app:eq:boost-operator:approx} in terms of expectation values of real\hyp{} and complex\hyp time evolved states, only:
\begin{align}
  G_{AB}(\omega)
  &\approx
  \int_0^T \mathrm dt\, \sum_{ij=1}^r \braket{\varphi_A|\hat U(t,\omega+\mathrm i\eta)|\psi^i_B}
  \sum_{\nu=1}^r Q_{i\nu} \frac{1}{1- \mathrm e^{-\mathrm i(E^\mathrm{eff}_\nu - E_0 - \omega - \mathrm i\eta)T}} \bar{Q}_{\nu j}
  \braket{\psi^j_B|\varphi_B} \\
  &=
  \int_0^T \mathrm dt\, \sum_{ij=1}^r
  \sum_{n=0}^{D-1} Y_{AB,n}(t,\omega+\mathrm i\eta) X_{ni}
  \sum_{\nu=1}^r Q_{i\nu} \frac{1}{1- \mathrm e^{-\mathrm i(E^\mathrm{eff}_\nu - E_0 - \omega - \mathrm i\eta)T}} \bar{Q}_{\nu j}
  \sum_{m=0}^{D-1} \bar{X}_{mj} M_{m0} \\
  &=
  \int_0^T \mathrm dt\, \vec{Y}_{AB}(t,\omega+\mathrm i\eta) 
  \mathbf X \mathbf Q \frac{1}{1- \mathrm e^{-\mathrm i(\mathbf E^\mathrm{eff} - E_0 - \omega - \mathrm i\eta)T}} \mathbf Q^\dagger \mathbf X^\dagger
  \vec{M}_0 \;. \label{app:eq:gf:final}
\end{align}
Note that in this expression all explicit dependencies on the basis states of the complex\hyp time Krylov subspace have been removed and the whole expression can be evaluated efficiently from computing the set of overlaps $M_{ij}, H^\mathrm{eff}_{ij}$ and $Y_{AB,n}(t)$, as well as a foremost evaluation of the contractions $\mathbf X \mathbf Q \frac{1}{1- \mathrm e^{-\mathrm i(\mathbf E^\mathrm{eff} - E_0 - \omega - \mathrm i\eta)T}} \mathbf Q^\dagger \mathbf X^\dagger
\vec{M}_0$.
The correction term $K^T_{AB}$ introduced in the main text is then easily extracted by separating the $0$th order from the expansion of the operator inverse and using the identity $\vec{Y}_{AB}(t,\omega+\mathrm i\eta) \mathbf X \mathbf Q \mathbf Q^\dagger \mathbf X^\dagger \vec{M}_0 = \vec{Y}_{AB}(t,\omega+\mathrm i\eta) \mathbf X \mathbf X^\dagger \vec{M}_0 = \braket{\varphi_A|\hat U(t,\omega+\mathrm i\eta)|\varphi_B}$:
\begin{equation}
  G_{AB}(\omega)
  =
  \int_0^T \mathrm dt\, \braket{\varphi_A|\hat U(t,\omega+\mathrm i\eta)|\varphi_B}
  +
  \int_0^T \mathrm dt\, \vec{Y}_{AB}(t,\omega+\mathrm i\eta) 
  \mathbf X \mathbf Q \frac{\mathrm e^{-\mathrm i(\mathbf E^\mathrm{eff} - E_0 - \omega - \mathrm i\eta)T}}{1- \mathrm e^{-\mathrm i(\mathbf E^\mathrm{eff} - E_0 - \omega - \mathrm i\eta)T}} \mathbf Q^\dagger \mathbf X^\dagger
  \vec{M}_0
  \equiv
  \tilde G^T_{AB} + K^T_{AB} \; .
\end{equation}
Some remarks are in order concerning the construction of the basis transformation $\mathbf X$ and the effective Hamiltonian $\mathbf H^\mathrm{eff}$.
In principle, the number $r$ of linearly independent basis states $\ket{\psi^j_\mu}$ is fixed by the rank of the Gram matrix $\mathbf M$.
However, due to finite precision arithmetics, rank\hyp revealing matrix factorizations such as the~\gls{SVD} only exhibit a finite resolution, which must be accounted for.
Usually, this is done by introducing a precision threshold $\epsilon$ which is of the order of the square root of the precision of the underlying arithmetics, typically $\epsilon = \sqrt{2^{-16}} \approx 10^{-8}$.
This can drastically reduce the dimension of the Krylov subspace and, therefore, deteriorate the approximation quality of the Green's function, especially if $\tau$ is chosen rather small.
In order to extract as many basis states as possible from the set of complex\hyp time evolved states, we therefore found it convenient to exploit the recursive~\gls{SVD} factorization suggested in~\cite{Stoudenmire:2013eq}.
Using this strategy, we could achieve precision thresholds $\epsilon = D \times 2^{-16}$, where $D$ is the dimension of the Gram matrix.
As an example, we may consider the computation of the Green's functions of the~\gls{SSH} model from the main text.
There, we chose a complex\hyp time evolution step size $\delta t=0.1$ yielding $D=321$ for a maximum simulation time $\mathrm {Re}(T^\mathrm{cplx})=32$.
The resulting precision threshold $\epsilon \approx 10^{-14}$ was small enough to ensure Krylov subspace dimensions $r\sim 300$ for all considered Green's functions.
Furthermore, we found it convenient to construct the basis transformation as $\mathbf X = \mathbf U \mathbf D^{-\nicefrac{1}{2}}$.
This way, we are directly working in the orthogonal basis, such that $\operatorname{dim} \mathbf H^\mathrm{eff} = r \leq D$, which would not be the case when transforming the coefficients of the orthonormalized basis states back into the basis of the complex\hyp time evolved states $\ket{\varphi^j_\mu}$.
\section{Error analysis\label{app:sec:error-analysis}}
In the derivation presented in~\cref{app:sec:derivation}, two relevant approximations were made, which we will discuss in the following.
First, the action of the boost operator was approximated in the complex\hyp time Krylov subspace
\begin{equation}
  \hat S(T,\omega+\mathrm i\eta) \ket{\varphi_B}
  \rightarrow
  \hat P^{D\tau}_B \hat S(T,\omega+\mathrm i\eta) \hat P^{D\tau}_B \ket{\varphi_B} \; . \label{app:eq:approx:boost-operator}
\end{equation}
Second, the boost operator itself was approximated within that very subspace by replacing the operator inverse
\begin{equation}
  \hat P^{D\tau}_B \left[1 - \hat U(T,\omega+\mathrm i\eta) \right]^{-1}\hat P^{D\tau}_B
  \rightarrow
  \left[ 1 - \mathrm e^{-\mathrm i(\hat P^{D\tau}_B \hat H \hat P^{D\tau}_B - E_0 - \omega - \mathrm i\eta)T} \right]^{-1} \; . \label{app:eq:approx:inverse}
\end{equation}
It is now convenient to reformulate the application of the boost operator as a linear algebra problem, remembering that for the case of~\gls{TN} methods, we are always working on a finite, yet exponentially large, Hilbert space.
For that purpose, we define $\mathbf H$ to be the matrix representation of the Hamiltonian, $\vec{\varphi}$ the vector representation of $\ket{\varphi_B}$ and $\vec \zeta$ the vector representation of the desired state $\hat S(T,\omega+\mathrm i\eta)\ket{\varphi_B}$, each of which being expanded in the same basis.
Then, $\vec \zeta$ is the solution to the system of linear equations
\begin{equation}
  \left[ 1-\mathrm e^{-\mathrm i(\mathbf H - E_0 - \omega - \mathrm i\eta)T} \right] \vec\zeta = \vec \varphi \; . \label{app:eq:boost-operator:lse}
\end{equation}
Denoting the matrix representation of the projector $\hat P^{D\tau}_B$ into the Krylov subspace by $\mathbf P$, the first approximation~\cref{app:eq:approx:boost-operator} can be written as
\begin{equation}
  \mathbf P \left[ 1-\mathrm e^{-\mathrm i(\mathbf H - E_0 - \omega - \mathrm i\eta)T} \right] \mathbf P \vec\zeta = \mathbf P\vec\varphi \; .
\end{equation}
This immediately shows that the second approximation~\cref{app:eq:approx:inverse} can be implemented as
\begin{equation}
  \left[ 1-\mathrm e^{-\mathrm i(\mathbf H^\mathrm{eff} - E_0 - \omega - \mathrm i\eta)T} \right] \vec\zeta = \mathbf P\vec\varphi \; , \quad \mathbf H^\mathrm{eff} = \mathbf P\mathbf H \mathbf P \; . \label{app:eq:approx}
\end{equation}
\subsection{Error bounds for the action of $\mathrm e^{-\mathrm i(\mathbf H - E_0 - \omega - \mathrm i\eta)T}$ in the complex\hyp time Krylov subspace\label{app:sec:error-analysis:hamiltonian}}
Let us begin by bounding the error in the second approximation~\cref{app:eq:approx}.
We abbreviate $\mathbf A = \mathbf H - E_0 - \omega - \mathrm i\eta$ and $\mathbf B_\tau = \mathrm e^{-\mathrm i \mathbf A \tau}$.
Note that in this section we use $\mathbf B_\tau$ as generator of the Krylov space, i.e., we replace $\mathbf H \rightarrow \mathbf H - E_0 - \omega - \mathrm i\eta$ in $\mathbf U_\tau$.
This transforms the spectrum of the generator to lie on a contracted circle $\sigma(\mathbf B_\tau) \cong \sigma(\mathbf U_\tau)\mathrm e^{-\eta\operatorname{Re}(\tau)}$ and rotates the location of the eigenvalues on that circle.
The rotations itself are of now further concern for the upcoming estimations, because they only constitute phase factors.
However, the contraction $-\eta\operatorname{Re}(\tau)$ is always evaluated in the limit $\eta\rightarrow 0$ when estimations containing the generator of the Krylov space are considered.
Using $T = N \delta t$ as well as $\mathbf A_\mathrm{eff} = \mathbf P \mathbf A \mathbf P$, we are interested in estimating
\begin{equation}
  \epsilon = \lVert \mathrm e^{-\mathrm i \mathbf A T} \vec\zeta - \mathrm e^{-\mathrm i \mathbf A_\mathrm{eff} T} \vec\zeta \rVert \; .
\end{equation}
The starting point is to use the Arnoldi relation for $\mathbf B_{\delta t - \delta\tau}$ and to note that
\begin{equation}
  \epsilon
  \leq
  T \mathrm e^{-\eta T} \lVert (\mathbf A - \mathbf P \mathbf A \mathbf P)\vec\zeta \rVert
  \leq
  2\mathrm e^{-\eta T} T \lVert (1 - \mathbf P)\mathbf A \mathbf P\rVert \cdot \lVert \mathbf P \vec\zeta \rVert \;,
\end{equation}
using Duhamel's integral formula for matrices $\mathrm e^{\mathbf S} - \mathrm e^{\mathbf T} = \int_0^1 \mathrm du\, \mathrm e^{(1-u)\mathbf S}(\mathbf S - \mathbf T)\mathrm e^{u\mathbf T}$.
Now we exploit the fact that in general $\eta>0$ is chosen small as possible such that we may assume $\Vert \mathbf B_{\delta t-\delta\tau} - 1\rVert < 1$.
Then the power series for the logarithm can be applied to write the expansion of $\mathbf A$ in the Krylov subspace as
\begin{equation}
  \mathbf A = \frac{\mathrm i}{\delta t - \delta\tau} \log(\mathbf B_{\delta t - \delta\tau}) = \frac{\mathrm i}{\delta t - \delta\tau} \sum_{n=1}^\infty (-1)^{n+1}\frac{(\mathbf B_{\delta t - \delta\tau} - 1)^n}{n}\;.
\end{equation}
The summands can be bound via
\begin{align}
  \lVert (1-\mathbf P)(\mathbf B_{\delta t - \delta\tau} - 1)^n \mathbf P\rVert
  &\leq 
  \lVert (\mathbf B_{\delta t - \delta\tau} - 1)^{n-1}\rVert \cdot \lVert (1-\mathbf P)(\mathbf B_{\delta t - \delta\tau} - 1) \rVert
  \leq
  \lVert \mathbf B_{\delta t - \delta\tau} - 1\rVert^{n-1} h_{r+1,r} \\
  \Rightarrow
  \lVert (1 - \mathbf P)\mathbf A \mathbf P\rVert
  &\leq
  \frac{1}{\lvert\delta t - \delta\tau\rvert}\sum_{n=1}^\infty \frac{\lVert (1-\mathbf P)(\mathbf B_{\delta t - \delta\tau} - 1)^n \mathbf P\rVert}{n}
  \leq
  \frac{N}{T\sqrt{1+\tan^2(\alpha)}} h_{r+1,r} \varphi(s) \; ,
\end{align}
where $\varphi(s) = -\log(1-s)/s$, $s = \lVert \mathbf B_{\delta t - \delta\tau} - 1 \rVert$, and $h_{r+1,r}$ denotes the Arnoldi residual for the $r$\hyp dimensional Krylov subspace.
We arrive at
\begin{equation}
  \epsilon
  \leq
  2N \frac{\mathrm e^{-\eta T}}{\sqrt{1+\tan^2(\alpha)}} h_{r+1,r} \varphi(s) \lVert \mathbf P \vec\zeta \rVert \;, \label{app:eq:error:A:arnoldi}
\end{equation}
which contains an explicit dependency on the complex angle as well as the Arnoldi residual.
Note that in particular the latter is expected to decay rapidly for large complex\hyp time Krylov space dimensions $r$, especially if $\mathrm e^{-\mathrm i \mathbf A T}$ acts on states with large contribution from the low\hyp energy subspace.
This statement can be made more precise by considering states $\vec \varphi = \sum_k c_k \vec \lambda_k$ with weights $c_k$ being centered around eigenstates $\vec\lambda_k$ of $\mathbf A$ with small eigenvalues $\lvert\lambda_k\rvert$.
The Arnoldi residual $h_{r+1,r}$ can be bounded by finding the degree $r-1$ polynomial $p(z)\in\mathbb P_{r-1}$ that best approximates $\mathbf B_{\delta t - \delta\tau}^r \vec\varphi$.
We use a decomposition $p(z) = \sum_k c_k q(z)$ ($q(z) \in \mathbb P_{r-1}$) such that, using $\gamma = \delta t - \delta\tau$, we can write the Arnoldi residual as
\begin{equation}
  h_{r+1,r}
  \simeq
  \inf_{p\in\mathbb P_{r-1}} \left\lVert \sum_k \left(c_k \mathrm e^{-\mathrm i \gamma \lambda_k r} - p(\mathrm e^{-\mathrm i \gamma \lambda_k})\right)\vec\lambda_k \right\rVert
  \leq
  \min_{p\in\mathbb P_{r-1}}\left(\sum_k \lvert c_k\rvert^2 \lvert \mathrm e^{-\mathrm i \gamma \lambda_k r} - q(\mathrm e^{-\mathrm i \gamma \lambda_k}) \rvert^2 \right)^{1/2} \;.
\end{equation}
Let us denote by $\tilde\lambda\in\mathbb R$ a cutoff eigenvalue separating the eigenvalues $\sigma(\mathbf A) \in [0, W-\mathrm i\eta] $ of $\mathbf A$ into two sets $\mathcal I_\leq = \left\{\lambda_k \in\sigma(\mathbf A) \vert\, \lvert\lambda_k\rvert \leq \tilde \lambda \right\}$ and $\mathcal I_> = \left\{\lambda_k \in\sigma(\mathbf A) \vert\, \lvert\lambda_k\rvert > \tilde \lambda \right\}$.
Then
\begin{equation}
  h_{r+1,r}
  \leq
  \left(\sum_{k\in\mathcal I_\leq} \lvert c_k\rvert^2 \lvert \mathrm e^{-\mathrm i \gamma \lambda_k r} - q(\mathrm e^{-\mathrm i \gamma \lambda_k}) \rvert^2 \right)
  +
  \left(\sum_{k\in\mathcal I_>} \lvert c_k\rvert^2 \lvert \mathrm e^{-\mathrm i \gamma \lambda_k r} - q(\mathrm e^{-\mathrm i \gamma \lambda_k}) \rvert^2 \right)
  \equiv
  R_\leq[q] + R_>[q]
\end{equation}
To obtain an explicit estimate we choose $q(z)$ to be the degree $r-1$ Taylor polynomial $q_{r-1}(z)$ of the function $z^r$, which provides a reasonably tight bound if we consider contributions from small eigenvalues of $\mathbf A$, since $q(z) \approx 1 + z$ if $\lvert z-1\rvert \ll 1$.
Using standard tail estimates for Taylor series we can bound $R_\leq[q_{r-1}]$
\begin{align}
  \max_{k\in\mathcal I_\geq} \lvert \mathrm e^{-\mathrm i \gamma \lambda_k r} - q_{r-1}(\mathrm e^{-\mathrm i \gamma \lambda_k}) \rvert
  &\leq
  \mathrm e^{ \operatorname{Im}(\gamma) \tilde\lambda r} \frac{(\lvert\gamma\rvert\tilde\lambda r)^r}{r!} \\
  \Rightarrow
  R_\leq[q_{r-1}]
  &\leq
  \underbrace{\sqrt{\sum_{k\in\mathcal I_\leq}\lvert c_k \rvert^2}}_{C_\leq} \mathrm e^{\operatorname{Im}(\gamma) \tilde\lambda r} \frac{(\lvert\gamma\rvert\tilde\lambda r)^r}{r!} \; , \label{app:eq:error:A:arnoldi:residual:lower}
\end{align}
where $C_\leq$ denotes the spectral weight of $\vec\varphi$ in the set of eigenstates $\left\{\lambda_k\right\}_{k\in\mathcal I_\leq}$.
For the contribution stemming from eigenvalues in $\mathcal I_>$ it suffices to establish a loose bound for the norm of the polynomials
\begin{align}
  \max_{k\in\mathcal I_>} \lvert \mathrm e^{-\mathrm i \gamma \lambda_k r} - q_{r-1}(\mathrm e^{-\mathrm i \gamma \lambda_k}) \rvert
  &\leq
  \mathrm e^{\operatorname{Im}(\gamma) \tilde\lambda r} + \max_{\lvert z\rvert \leq \mathrm e^{\operatorname{Im}(\gamma) \tilde\lambda r}} \lvert q(z) \rvert
  \equiv
  D_> \\
  \Rightarrow
  R_>[q_{r-1}]
  &\leq
  C_> \cdot D_> \;, \label{app:eq:error:A:arnoldi:residual:upper}
\end{align}
where we defined $C_> = \sqrt{\sum_{k\in\mathcal I_>}\lvert c_k \rvert^2}$ similar to $C_\leq$.
Summing up both contributions and using $\lvert \gamma \rvert = \sqrt{1+\tan^2(\alpha)} T/N$ we arrive at the following error bound for the Arnoldi residual
\begin{equation}
  h_{r+1,r}
  \leq
  C_\leq \cdot \mathrm e^{-\tan(\alpha)T\tilde\lambda r/N} \frac{\left(\sqrt{1+\tan^2(\alpha)}T\tilde\lambda r/N\right)^r}{r!} + C_> \cdot D_> \label{app:eq:error:A:arnoldi:residual}
\end{equation}
In the previous calculations we repeatedly used the relation $\lvert \mathrm e^{-\mathrm i \gamma s}\rvert \leq \mathrm e^{\operatorname{Im}(\gamma)s} = \mathrm e^{-T\tan(\alpha)s/N}$ for $s>\mathbb R^+_0$.
The resulting error bound for $h_{r+1,r}$ illustrates the crucial relevance of the angle $\alpha$ specifying the complex\hyp time contour used to generate the Krylov subspace.
In fact, as soon as $\alpha > 0$, the constant $D_>$ determining the leakage into the large eigenvalue subspace of $\mathbf A$ exhibits an exponential decay with the Krylov subspace dimension $r$.
This allows to obtain highly precise approximations of the small eigenvalue subspace even if the spectral weight $C_>$ is not negligible, and in stark contrast to the case where $\alpha = 0$.
This is a crucial feature rendering complex\hyp time Krylov space an ideal candidate to study for instance critical systems.
Nevertheless, care must be taken due to the presence of the $\varphi(s)$\hyp dependence in~\cref{app:eq:error:A:arnoldi}, prohibiting too large values of $\alpha$.
\subsection{Error bounds for the action of $[1-\mathrm e^{-\mathrm i(\mathbf H - E_0 - \omega - \mathrm i\eta)T}]^{-1}$ in the complex\hyp time Krylov subspace\label{app:sec:error-analysis:boost-operator}}
We now turn to an analysis of the approximation errors occuring when solving~\cref{app:eq:boost-operator:lse} in the complex\hyp time Krylov subspace.
Denote by $\vec\zeta_r$ the solution of~\cref{app:eq:boost-operator:lse} projected in the $r$\hyp dimensional complex\hyp time Krylov subspace and replacing $\mathbf A \rightarrow \mathbf A_\mathrm{eff}$.
The approximation error is given by
\begin{align}
  {\vec e}_{\vec \varphi}
  &=
  \vec\zeta - \vec\zeta_r = \left[(1-\mathrm e^{-\mathrm i\mathbf A T})^{-1} - (1-\mathrm e^{-\mathrm i\mathbf A_\mathrm{eff} T})^{-1} \right]\vec\varphi \\
  \Rightarrow
  \epsilon_{\vec\varphi} := \lVert {\vec e}_{\vec\varphi} \rVert
  &\leq
  \lVert
    (1-\mathrm e^{-\mathrm i\mathbf A T})^{-1}
  \rVert
  \cdot
  \lVert
    \mathrm e^{-\mathrm i\mathbf A T} - \mathrm e^{-\mathrm i\mathbf A_\mathrm{eff} T}
  \rVert
  \cdot
  \lVert
    (1-\mathrm e^{-\mathrm i\mathbf A_\mathrm{eff} T})^{-1}
  \rVert
  \cdot
  \lVert \vec\varphi \rVert \; ,
\end{align}
where we used for two matrices $\mathbf T, \mathbf R$ the resolvent identity: $(1-\mathbf T)^{-1} - (1-\mathbf R)^{-1} = (1-\mathbf T)^{-1}(\mathbf T - \mathbf R)(1-\mathbf R)^{-1}$.
The first and last matrix norm can be bounded directly
\begin{equation}
  \lVert
    (1-\mathrm e^{-\mathrm i\mathbf A T})^{-1}
  \rVert
  \cdot
  \lVert
    (1-\mathrm e^{-\mathrm i\mathbf A_\mathrm{eff} T})^{-1}
  \rVert
  \leq
  \frac{1}{(1-\mathrm e^{-\eta T})^2} \;,
\end{equation}
using the fact that the eigenvalues of $\mathbf A,\mathbf A_\mathrm{eff}$ are of the form $\lambda_k = \varepsilon_k -\mathrm i\eta$ with $\varepsilon_k,\eta \in \mathbb R$ and $\eta>0$.
A bound for the intermediate matrix norm has already been derived and is given by~\cref{app:eq:error:A:arnoldi}.
We arrive at the following bound for the approximation error of the action of the boost operator
\begin{equation}
  \epsilon_{\vec\varphi}
  \leq
  \frac{2N}{\sqrt{1+\tan^2(\alpha)}} \frac{e^{-\eta T}}{(1-\mathrm e^{-\eta T})^2} h_{r+1,r} \varphi(s) \;, \label{app:eq:error:S:residual}
\end{equation}
where the Arnoldi residual $h_{r+1,r}$ is given by~\cref{app:eq:error:A:arnoldi:residual} and we used $\lVert\vec\varphi\rVert = 1$.
We close this section by simplifying the bound employing practically relevant estimations for some of the parameters occuring in~\cref{app:eq:error:A:arnoldi:residual,app:eq:error:A:arnoldi:residual}.
First we note that the standard frequency resolution of spectral functions evaluated from time\hyp evolutions is given by the Nyquist\hyp Shannon limit and implies $\eta T \sim \mathcal O(1)$.
We thus are interested in a regime where $\eta T < 1$ or even $\eta T \ll 1$, in which we can approximate $\mathrm e^{-\eta T}/(1-\mathrm e^{-\eta T})^2 \approx 1/(\eta T)^2$.
Note that without further regularization this asymptotic behavior would prohibit a significant increase in the low\hyp energy precision.
We can also simplify the residual making use of the fact that for the complex\hyp time Krylov space large dimensions $\sim \mathcal O(100) - \mathcal O(1000)$ are possible, as well as $\alpha \sim \mathcal O(0.01)$.
This permits to employ Stirling's formula to approximate
\begin{equation}
  \frac{\left(\sqrt{1+\tan^2(\alpha)}T\tilde\lambda r/N\right)^r}{r!}
  \approx
  \frac{\left(\mathrm e T\tilde\lambda/N\right)^r}{\sqrt{2\pi r}}
\end{equation}
Combining these approximations with $\varphi(\lVert \mathbf B_\gamma -1\rVert) \approx 1$ we obtain the relevant error approximation for applying the boost operator to some initial guess state
\begin{equation}
  \epsilon_{\vec\varphi}
  \lesssim
  \frac{2N}{(\eta T)^2} \frac{\left(\mathrm e T\tilde\lambda/N\right)^r}{\sqrt{2\pi r}} \mathrm e^{-\tan(\alpha)T\tilde\lambda r/N}
  =
  \frac{2}{\delta t T \eta^2} \frac{\mathrm e^{r (1 + \ln(\delta t \tilde\lambda) - \tan(\alpha) \delta t\tilde\lambda)}}{\sqrt{2\pi r}} \; ,
\end{equation}
where $\tilde\lambda$ denotes a cutoff eigenvalue above which $\vec\varphi$ has vanishing spectral weight and we reinserted the complex\hyp time evolution step size $\delta t = T/N$.
Several observations can be made, for which we now return to a Krylov space generated by $\mathrm e^{-\mathrm i\mathbf H (1-\mathrm i\tan(\alpha))\delta t}$.
First of all, as long as $\delta t \tilde\lambda < 1/\mathrm e$ we have exponential convergence independent on the complex angle $\alpha$.
This appears to be a general feature of Krylov space approximations of low\hyp lying eigenvalues build from time evolutions and could help to analyze Krylov\hyp space based methods for state preparation on quantum computers\cite{Shen2023}.
Second, for finite $\alpha$ the range of exponential convergence is determined by the solutions to the equation
\begin{equation}
  h(\delta t\tilde\lambda) := 1+\ln(\delta t\tilde \lambda) - \tan(\alpha)\delta t\tilde\lambda \stackrel{!}{=} 0 \;,
\end{equation}
which come in pairs, if $0<\tan(\alpha) < 1/\mathrm e$, otherwise there is always exponential convergence for any $\tilde\lambda$.

Assuming now $0 < \tan(\alpha) < 1/\mathrm e$, then for fixed $\delta t$ there are $0 < \tilde\lambda_{c,1} \leq \tilde\lambda_{c,2}$ such that $h(\lambda < \tilde\lambda_{c,1}) < 0$ and $h(\lambda > \tilde\lambda_{c,2}) < 0$.
Therefore, the complex angle increases the low\hyp energy range of exponential convergence and furthermore introduces a second, high\hyp energy regime in which the application of the boost operator is again faithfully captured.
Finally, the location $\tilde\lambda_{c,1}$ can be pushed to larger eigenvalues by reducing the step size $\delta t$.
This provides an additional, computationally subleading leverage to increase the energy range captured by the complex\hyp time Krylov space.
\subsection{Error bound for the Green's function augmented by a complex\hyp time Krylov subspace\label{app:sec:error-analysis:greens-function}}
We are now in the position to formulate an error bound for the Green's function when expanding the boost operator in the complex\hyp time Krylov subspace.
We denote by $G^K_{AB}(\omega+\mathrm i\eta)$ the Green's function obtained when approximating the boost operator in the complex\hyp Krylov space and consider
\begin{align}
  \delta G^K_{AB}(\omega+\mathrm i\eta)
  &=
  \int_0^T \mathrm dt\, \bra{\varphi_A} \hat U(t,\omega+\mathrm i\eta) \left(\frac{1}{1-\mathrm e^{-\mathrm i (\hat H-E_0-\omega-\mathrm i\eta)T}} - \frac{1}{1-\mathrm e^{-\mathrm i (\hat H_\mathrm{eff}-E_0-\omega-\mathrm i\eta)T}} \right) \ket{\varphi_B } \\
  \Rightarrow
  \lvert \delta G^K_{AB}(\omega+\mathrm i\eta) \rvert
  &\leq
  \frac{1-\mathrm e^{-\eta T}}{\eta}\lVert (1-\mathrm e^{-\mathrm i\mathbf A T})^{-1} - (1-\mathrm e^{-\mathrm i\mathbf A_\mathrm{eff} T})^{-1} \rVert \notag \\
  &=
  \frac{e^{-\eta T}}{\eta(1-\mathrm e^{-\eta T})} \frac{2N}{\sqrt{1+\tan^2(\alpha)}} h_{r+1,r} \varphi(s) \;, \label{app:eq:error:G:ctks}
\end{align}
where we used~\cref{app:eq:error:S:residual} to bound the approximation error in the boost operator for $\omega\in[0,W]$.
The resulting error estimation differs from that for the boost operator only by a prefactor $\frac{1-\mathrm e^{-\eta T}}{\eta}$ such that the asymptotic dependency of $\delta G^K_{AB}(\omega+\mathrm i\eta)$ on $\eta,T$ for $\eta T \ll 1$ is $1/\eta^2 T$.
We thus arrive at the relevant error bound for the complex\hyp time Krylov subspace augmented Green's function
\begin{equation}
  \delta G^K_{AB}(\omega+\mathrm i\eta)
  \lesssim
  \frac{2}{\delta t \eta^2} \left(\frac{\mathrm e^{r (1 + \ln(\delta t \tilde\lambda) - \tan(\alpha) \delta t\tilde\lambda)}}{\sqrt{2\pi r}} + C_> D_> \right) \; \label{app:eq:error:G:small-etaT},
\end{equation}
where again $r$ denotes the dimension of the complex\hyp time Krylov subspace, $\delta t = T/N$ and $\tilde\lambda$ is a cutoff eigenvalue of $\hat H$, which, for convenience, can be chosen such that $\ln(\delta t \tilde\lambda) < -1$.
Note that from~\cref{app:eq:error:A:arnoldi:residual:upper} the dominating contribution $D_>$ of the leakage into the high energy subspace is exponentially suppressed in the Krylov subspace dimension owing to the finite complex angle $\alpha$.
We emphasize that this suppresion is absent when using a real\hyp time evolution only.
We can compare the derived error bound to the error incurred when using only the naive Fourier transform on the finite time interval $[0,T]$ to determine the Green's function, we will refer to the latter as $G^T_{AB}(\omega+\mathrm i\eta)$.
Using the same estimations for the integration of the time evolution operator but replacing $\mathrm e^{-\mathrm i\mathbf A_\mathrm{eff}T} \rightarrow 1$ we obtain
\begin{equation}
  \lvert \delta G^T_{AB}(\omega+\mathrm i\eta) \rvert
  \leq
  \frac{1-\mathrm e^{-\eta T}}{\eta}\underbrace{\lVert (1-\mathrm e^{-\mathrm i\mathbf A T})^{-1} - 1 \rVert}_{\leq \mathrm e^{-\eta T}/(1-\mathrm e^{-\eta T})}
  \leq
  \frac{\mathrm e^{-\eta T}}{\eta} \;. \label{app:eq:error:G:naive}
\end{equation}
Comparing~\cref{app:eq:error:G:ctks,app:eq:error:G:naive} then allows to use the Arnoldi residual of $\mathbf H$ to derive an expression for the parameter regime, in which the complex\hyp time Krylov space approximation can be expected to yield better results
\begin{align}
  \delta G^K_{AB}(\omega+\mathrm i\eta)
  &\stackrel{!}{<}
  \delta G^T_{AB}(\omega+\mathrm i\eta) \notag \\
  \Leftrightarrow
  h_{r+1,r}
  &<
  \frac{1-\mathrm e^{-\eta T}}{2N} \quad
  \stackrel{\eta T \ll 1}{\Rightarrow}
  h_{r+1,r} < \frac{\eta\delta t}{2} \label{app:eq:error:G:comparison}
\end{align}
where again we approximated $\sqrt{1+\tan^2(\alpha)} \approx 1$ and $\varphi(s)\approx 1$ for small enough $\alpha$.
Employing Stirling's formula to again rewrite the Arnoldi residual for large Krylov space dimensions we obtain a more explicit, yet due to the presence of the cutoff eigenvalue $\tilde\lambda$ numerically less handy, form
\begin{equation}
  \frac{\mathrm e^{r (1 + \ln(\delta t \tilde\lambda) - \tan(\alpha) \delta t\tilde\lambda)}}{\sqrt{r}} < \sqrt{\frac{\pi}{2}} \eta\delta t \;.
\end{equation}
Here, we neglected the high\hyp energy leakage $C_>D_>$, which requires $r\ln(\delta t\tilde\lambda)\ll -1$, i.e., the Krylov space dimension must be large enough to compensate a possibly slow decay of the spectral decomposition of $\ket{\varphi_B}$.
We close this error analysis with some comments on the utility of the derived error bounds as well as their impact on actual simulations.
At first we note that for practical purposes evaluating the Arnoldi residual should be prefered, because once $\mathbf H_\mathrm{eff}\in\mathbb C^{r+1\times r+1}$ is known, it can be brought easily into upper Hessenberg form to read of $h_{r+1,r}$.
With $h_{r+1,r}$ at hand we can directly evaluate~\cref{app:eq:error:G:ctks} for given $\eta, T$ as well as~\cref{app:eq:error:G:comparison} to decide whether the convergence of the Krylov space representation is sufficient.
From~\cref{app:eq:error:G:comparison} we also observe that due to the exponential suppression of the Arnoldi error with the time step size $\delta t$, it should always be attempted to rather increase the frequency of complex\hyp time Krylov space steps, instead of reaching large values of the final evolution time $T$.
This can be accompanied by tuning the complex angle $\alpha$ as large as possible, i.e., $\alpha \sim 0.1$, while keeping in mind that for the evolved time frame the states $\hat U^D_\tau\ket{\varphi_B}$ must not be linearly dependent.
The latter can be tested easily by monitoring the rank of the Gram matrix $\mathbf M$ of the evolved states.
In computationally challenging situations, the bound ~\cref{app:eq:error:G:small-etaT} can still be too loose.
In those cases, we can employ the fact that errors are predominantly generated by mismatching poles, i.e., unconverged eigenvalues of the Hamiltonian in the complex time Krylov subspace.
This is easily seen by expanding the Green's function in the Lehmann representation and using the fact that for any finite broadening $\eta>0$ this yields a sum of Lorentzians of width $\eta$.
We can thus define the number of maximally resolvable peaks via $n = D/\eta$ in the spectral width $D$ of the Green's function to arrive at the mean error per peak
\begin{equation}
  R
  =
  \frac{\delta G^K_{AB}(\omega+\mathrm i\eta)}{n}
  \lesssim
  \frac{2}{\delta t \eta D} h_{r+1,r} \; .
\end{equation}
This provides an upper bound that can be used as a threshold to discriminate spectral signals from errors such as replica features.
\subsection{Numerical convergence of complex\hyp time Krylov subspace\label{app:sec:error-analysis:spectral-stability}}
\begin{figure}
  \centering
  \subfloat[\label{fig:ctks-convergence:x-1.0.y-0.0}]{
    \includegraphics[width=0.48\textwidth]{\figurepath/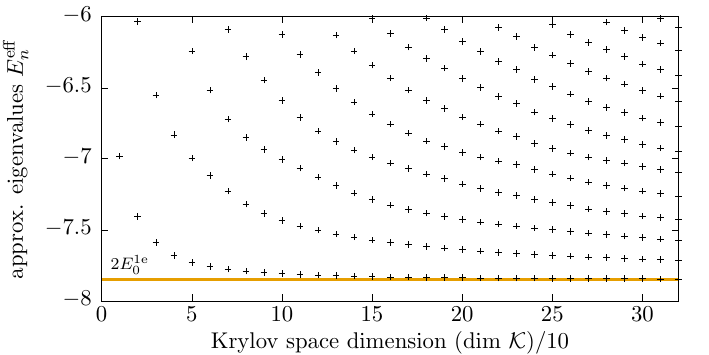}
  }\hfill%
  \subfloat[\label{fig:ctks-convergence:x-0.0.y-1.0.kx-0.0.ky-0.0}]{
    \includegraphics[width=0.48\textwidth]{\figurepath/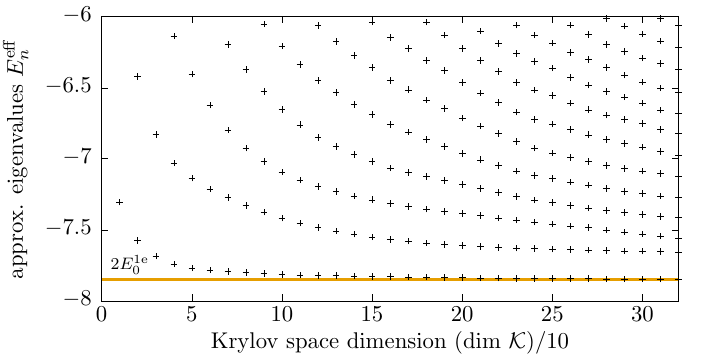}
  }
  \caption{
    Convergence of the low\hyp lying eigenstates of~\cref{eq:ssh-model} as a function of the Krylov subspace dimension on an $8\times 8$ square lattice.
    \subfigref{fig:ctks-convergence:x-1.0.y-0.0} shows eigenvalues obtained when constructing the Krylov subspace using~\cref{eq:initial-state:r-space} with $\vec a = (0,1)^t$ as initial state.
    \subfigref{fig:ctks-convergence:x-0.0.y-1.0.kx-0.0.ky-0.0} shows eigenvalues obtained when constructing the Krylov subspace using~\cref{eq:initial-state:k-space} as initial state with $\vec k = (0,0)^t$ and $\vec a = (0,1)^t$.
    Comparison with the two\hyp polaron energy $2E^\mathrm{1e}_0$ indicated by the yellow line demonstrates that the lowest eigenvalues are already well\hyp converged in both cases.
  }
\end{figure}
We performed a systematic convergence analysis of the complex\hyp time Krylov subspace representation by monitoring the approximated eigenvalues of the two\hyp dimensional~\gls{SSH} model
\begin{equation}
  \hat H
  =
  -\sum_{\langle \vec r, \vec r^\prime\rangle}\sum_{\sigma} \left(\hat c^\dagger_{\vec r,\sigma}\hat c^\nodagger_{\vec r^\prime,\sigma} + \mathrm{h.c.} \right) \left(t - g\hat V^\nodagger_{\vec r, \vec r^\prime} \right)
  +U\sum_{\vec r} \hat n^\nodagger_{\vec r,\uparrow} \hat n^\nodagger_{\vec r,\downarrow} + \Omega\sum_{\vec r} \hat a^\dagger_{\vec r} \hat a^\nodagger_{\vec r} \; , \label{eq:ssh-model}
\end{equation}
in the two\hyp particle sector on a square lattice of size $L\times L$ with $L=8$
Here, $\hat c^{(\dagger)}_{\vec r,\sigma}$ are the spin\hyp $\sigma$ electronic ladder operators, $\hat a^{(\dagger)}_{\vec r}$ denote the annihilation (creation) operators of optical phonons with frequency $\Omega$ and the sum $\sum_{\langle \vec r, \vec r^\prime\rangle}$ is over nearest neighboring lattice sites $\vec r, \vec r^\prime$.
The interaction is given by $\hat V_{\vec r, \vec r^\prime} = \sqrt{2 M \Omega} \left( \hat X^\nodagger_{\vec r} - \hat X^\nodagger_{\vec r^\prime} \right)$ with the displacements $\hat X^\nodagger_{\vec r}=\frac{1}{\sqrt{2M\Omega}}(\hat a^\dagger_{\vec r} + a^\nodagger_{\vec r})$, and we chose the oscillator mass $M = 1$ as well as $t=1$ as unit of energy and $\Omega/t=0.2$.
All subsequeny analysis is performed in the strong coupling regime $\lambda=\frac{g^2}{2\Omega t}=0.2$.
We constructed two types of initial states
\begin{align}
  \ket{\vec a} &= \hat c^\dagger_{\vec c,\uparrow}\hat c^\nodagger_{\vec c+\vec a, \downarrow} \ket{\varnothing} \label{eq:initial-state:r-space}\; , \\
  \ket{\vec k,\vec a} &= \frac{1}{L} \sum_{\vec r} \mathrm e^{-\mathrm i \vec k (2\vec r + \vec a)} \hat c^\dagger_{\vec r,\uparrow}\hat c^\nodagger_{\vec r+\vec a, \downarrow} \ket{\varnothing} \label{eq:initial-state:k-space}
\end{align}
where $\vec c = (L/2,L/2)^t$.
For illustrative purposes, we show the convergence of the low\hyp lying eigenstates of~\cref{eq:ssh-model} in complex\hyp time Krylov spaces constructed from~\cref{eq:initial-state:r-space} with $\vec a = (0,1)^t$ in~\cref{fig:ctks-convergence:x-1.0.y-0.0}, and from~\cref{eq:initial-state:k-space} with $k=\vec(0,0)^t$ and $\vec a = (0,1)^t$ in~\cref{fig:ctks-convergence:x-0.0.y-1.0.kx-0.0.ky-0.0}.
Also, we display the two\hyp polaron energy $2E^\mathrm{1e}_0$ indicating the onset of unbound two\hyp electron states.
We find rapid convergence of the lowest eigenvalue towards the two\hyp polaron energy consistent with our findings of the absence of a bipolaron\ signal in the spectral functions in the main text.
Note that the location of the first excited state, which is already well\hyp converged in the case of~\cref{eq:initial-state:k-space}, differs between the two initial states.
This can be a consequence of the fact that by construction~\cref{eq:initial-state:k-space} is orthogonal to two\hyp particle states with momenta $\vec q \neq \vec k$, thus suppressing spectral leakage into high energy subspaces as discussed in~\cref{app:sec:error-analysis:greens-function}.
\begin{figure}
  \centering
    \includegraphics[width=0.95\textwidth]{\figurepath/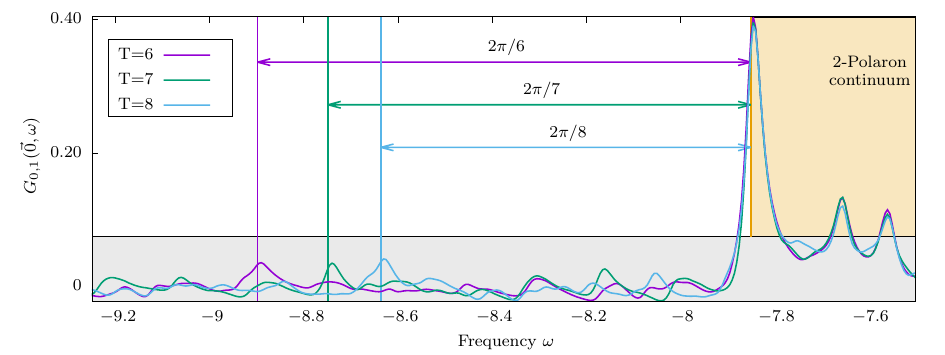}
  \caption{
    \label{fig:ctks-replica:x-0.0.y-1.0.kx-0.0.ky-0.0}
    Convergence of the $S1$ bipolaronic spectral function of~\cref{eq:ssh-model} when varying the real\hyp time evolution time $T$ on an $8\times 8$ square lattice.
    Constructing the Krylov subspace using~\cref{eq:initial-state:k-space} as initial state with $\vec k = (0,0)^t$ and $\vec a = (0,1)^t$, the effect of replica features from the two\hyp polaron energy $2E^\mathrm{1e}_0$ (yellow line) caused by unconverged higher many\hyp body eigenstates is illustrated.
    The position of these replica (or ringing) features is expected at multiples of the frequency resolution set by $2\pi/T$ and the marked positions clearly demonstrate the largest features below the error threshold (gray shaded area) are not of physical origin.
  }
\end{figure}
The convergence of the complex\hyp time Krylov subspace augmented spectral functions can also be checked by investigating the presence of replica feautres expressing themselves in form of oscillations with frequencies $2\pi/T$.
They are well\hyp known effects of finite time\hyp domain Fourier transforms and merely a consequence of the Nyquist\hyp Shannon sampling theorem.
Involving~\cref{app:eq:boost-operator:approx}, unconverged many\hyp body eigenstates result in an incomplete cancellation of these replica features generated by the real\hyp time evolution
Due to the presence of weak signals below the error threshold in the bipolaronic spectral functions, we checked whether these signals may by of physical origin or caused by unconverged many\hyp body eigenstates.
For that purpose, we varied the real\hyp time evolution time $T$ and computed the $S1$ bipolaron spectral functions using ~\cref{eq:initial-state:k-space} as initial state with $\vec k = (0,0)^t$ and $\vec a = (0,1)^t$.
The results are shown in~\cref{fig:ctks-replica:x-0.0.y-1.0.kx-0.0.ky-0.0} where we reduced the broadening by a factor of two compared to the main tex, i.e., $\eta=0.05$, to enhance the amplitude of the signals below the two\hyp polaron continuum.
We observe an excellent agreement for the expected positioning of the dominating signals when varying the real\hyp time evolution range between $T=6,7,8$.
This further supports both, the validity of the error estimation, and the absence of bipolarons described by~\cref{eq:initial-state:k-space} in the strong coupling regime.
\bibliography{literature}